\documentclass[conference]{IEEEtran}
\usepackage{comment}

\usepackage{blindtext}
\usepackage{amssymb}
\usepackage{subfigure}
\usepackage{amsmath}
\usepackage{amsfonts}
\usepackage{mathrsfs}
\usepackage{engord}
\usepackage[dvips]{graphicx}
\usepackage{bbm}
\usepackage{threeparttable}
\usepackage{booktabs}
\usepackage{epsfig}
\usepackage{color}
\usepackage{graphicx}
\usepackage{multirow}
\usepackage{cite}
\usepackage{epstopdf}
\usepackage{amsthm}
\usepackage{framed}
\usepackage{url}

\newtheorem{lem}{Lemma}

\usepackage[ruled,norelsize]{algorithm2e}

\makeatletter
\newcommand{\removelatexerror}{\let\@latex@error\@gobble}
\makeatother

\begin{document}

%
\title{Efficient, Effective and Well Justified  Estimation of Active Nodes within a Cluster}

\author{\IEEEauthorblockN{Md Mahmudul Hasan,
Shuangqing Wei, Ramachandran Vaidyanathan}}
\maketitle
\footnotetext[1]{Md M Hasan,  S. Wei and R. Vaidyanathan are with the school of Electrical Engineering and Computer Science, Louisiana State University, Baton Rouge, LA 70803, USA (Email: mhasa15@lsu.edu, swei@lsu.edu, vaidy@lsu.edu). }


\begin{abstract}
Reliable and efficient estimation of the size of a dynamically changing cluster in an IoT network is critical in its nominal operation.  Most previous estimation schemes worked with  relatively smaller frame size and large number of rounds. Here we  propose a new estimator named \textquotedblleft Gaussian Estimator of Active Nodes,\textquotedblright (GEAN), that works with large enough frame size  under which testing statistics is well approximated as a Gaussian variable,  thereby requiring less number of frames, and thus less total number of channel slots to attain a desired accuracy in estimation. More specifically, the selection of the frame size is done according to Triangular Array Central Limit Theorem  which also enables us to quantify the approximation error. Larger frame size helps the statistical average to converge faster to the ensemble mean of the estimator and the quantification of the approximation error helps to determine the number of rounds to keep up with the accuracy requirements. We present the analysis of our scheme under two different channel models i.e. $ \{0,1 \} $ and $ \{0,1,e \} $, whereas all previous schemes worked only under $ \{0,1 \} $ channel model. The overall performance of GEAN is  better than the previously proposed schemes considering the number of slots required for estimation to achieve a given level of estimation accuracy.   
\end{abstract}


\section{INTRODUCTION}
In a world pervaded with Internet of Things (IoT), small distance communications, now a days, carry more value than ever. A very common scenario is a set of agents communicating with a central access point (AP) and exchanging infomation. Typical examples would be some self powered chips attached to parked vehicles in a lot sending and receiving infomation from a common source, locally placed sensors accumulating information and sending to a base station or tags attached to the inventories  in a supermarket sending and receiving signal to and from a reader helping the inventory control \cite{nemmaluri2008sherlock}\cite{lee2008efficient}\cite{klaus2010rfid} \cite{zhao2019sensor}\cite{zhang2019heterogeneous} \cite{gurugopinath2019cache} \cite{zhang2019performance}. In all the above scenarios, knowing the number of active agents (AG) at a given time i.e. the nodes that are communicating with the AP, is of paramount importance. And that gives rise to a rich vein of reserch area that caters to solving the problem of estimating the number of active agents in the close vicinity of a common AP. In a very broad sense, such estimation calls for techniques that offer estimation reliability while ensuring minimum possible consumption of resources to do so. For a more concrete understanding of such estimation technicques and the respective reliability, the problem of estimating the number of active nodes communating to a central server  can be formally stated as follows: for a given reliability requirement $\alpha \in[0,1)$, a confidence interval $\beta \in [0,1)$ a central node will have to estimate an unknown population of active agents  $t$ in its vicinity. The estimation has to maintain the minimum accuracy condition $P[|\hat{t}-t|\leq  \beta t]\geq \alpha$ where $\hat{t}$ is the estimated value of the actual number of active active agents $t$. 

One of the early works in this trail was Unified Probabilistic Estimator (UPE) proposed in \cite{kodialam2006fast}. UPE estimation was based on the number of empty slots in a frame or the number of collision slots in the frame, where \textit{empty} slots are the slots that have not been replied to by any of the AGs, \textit{singleton} slots are the ones that have been replied to by exactly one of the AGs and \textit{collision} slots are the slots that have been replied to by more than one AGs. UPE has larger variance  which only meant more number of rounds required. An improved framed slotted Aloha protocol-based estimation called Enhanced Zero Based (EZB) estimator was proposed in \cite{kodialam2007anonymous}. EZB makes its estimation based on the total number of empty slots in a frame. The difference between EZB and UPE is, UPE makes an estimation of the population size in each frame and at the end averages out all the estimation results, whereas EZB finds the average of the number of $0$s in each frame and finally makes the estimation based on this average value.
First Non Empty Based (FNEB) estimator proposed in \cite{han2010counting} is based on the average number of slots before the first $ 1 $ appears in a frame .   Maximum Likelihood Estimator
(MLE) proposed in \cite{li2010energy} came with the motive to minimize power consumption by the AGs. The multireader  estimation proposed  in \cite{shah2009anonymous} assumes that any AG covered by several APs replies to only one of them. 
 Collision Set Estimator (CSE), proposed in \cite{zanella2012estimating}  uses maximum likelihood estimation to estimate the population size. CSE does not take accuracy requirements into account, hence cannot achieve required level of reliability. An algorithm to  estimate the cardinality or RFID tags under multiple readers with overlaping regions was proposed in \cite{shah2011cardinality}. ART \cite{shahzad2015fast} estimates the population size based on the average run size of $1$s in a frame where $1$ stands for a nonempty slot. For each frame ART calculates the average run length of $ 1 $s. After $ n $  such rounds ART uses the average of these averages to estimate the  population size by an invertible function.  A tag identification technique was proposed by the same authors in \cite{shahzad2013probabilistic}. In the two subsequent paragraphs we describe the major differences between our approach i.e. Gaussian Estimator of Active Nodes (GEAN) and the other existing schemes. 
 
 Firstly, all the other schemes worked with smaller frame sizes and relied on larger number of rounds to meet the accuracy requirements. Unlike them, we work with bigger frame size and smaller number of frames. With small frames sizes for any statistics to be approximated as Gaussian, the other schemes have to run large number of rounds. In contrast, GEAN selects a  large enough frame size according to Triangular Array Central Limit Theorem \cite{araujo1980central} to ensure that the estimator is Gaussian distributed within a frame. So, our scheme does not have to  rely on the number of rounds for its estimator to be approximated as Gaussian. The only reason that we need to play more than one round is to counter the impact of  variances of the estimator. Larger frame size helps the convergence speed of the  statistical average of the estimator to its ensemble mean. Another telling advantage a larger frame size gives is that, it decreases the probability  of a slot to have more than one reply and increases the probabiliy with which you can trust a slot to have either no reply or exactly one reply, which leads to better estimation accuracy.  
 
 Secondly, unlike the other works who naively assumed their estimators to be Gaussian distributed, we quantified the approximation error of our estimator within a frame to Gaussian distribution with all the necessary details. Any estimator approximated as Gaussian incurs some approximation error. It's important to quantify the approximation error to get the measure of the error you are making and how badly does that error impact the overall estimation accuracy. In this paper we have dedicated a subsection to talk about approximation error, its quantification and its impact in the overall estimation accuracy. 
 
 Precise selection of frame size along with accurate approximation of the estimator gives us advantages both in terms of saving the resources and achieving the desired estimation accuracy. Operationally speaking, such analysis give us advantages over other schemes both in terms of reliability and sparing use of resources. Some of the previously proposed schemes offers reliable estimation but uses up too many frame slots for estimation, where as some others are economical in terms of their use of the number of frame slots but fail to deliver on the promised reliability. We have presented relative comparison of the performance of GEAN with other works both in terms of reliability and respective usage of frame slots. And the results clearly shows that GEAN performs better than the other well known schemes if the overall package of reliability and usage of resources considered together. 
   
Unlike other schemes we have presented our scheme under two different channel models i.e. $\{ 0,1\}$ and $\{ 0,1,e\}$. Under  $ \{0,1\} $ channel model  each empty slot is represented by a $ 0 $ and each non-empty slot is represented by a $ 1 $, whereas under $ \{0,1,e\} $ channel model  each empty slot is represented by a $ 0 $, each singleton slot is represented by a $ 1 $ and each collision slot is represented by an $ e $. The motivation for $\{ 0,1,e\}$ is that, it achieves the desired accuracy in lesser number of time slots than $\{ 0,1\}$ channel models. This is because under $\{ 0,1,e\}$ each $ 1 $ is  definite whereas under $\{ 0,1\}$ a $ 1 $ can actually be exactly $ 1 $ or a number greater than $ 1 $. This added assurance helps $\{ 0,1,e\}$ model to achieve the desired accuracy using lesser number of slots, or equivalently gives better accuracy at the expense of the same number of slots . It is important to note that, $\{ 0,1,e\}$ channel model has the added cost of distinguishig between a $ 1 $ and an $ e $. For someone willing to pay more for a more accurate and assured estimation, $\{ 0,1,e\}$ is an automatic choice over $\{ 0,1\}$ channel model. Next we enumerate the prominent features of our appraoch. 
  \begin{enumerate}
   \item The scheme uses large enough frame size selected  according to Triangular Array Central Limit Theorem, that makes the estimator distribution within a frame a well justified Gaussian random variable.
   \item We have analytically quantified the approximation error of our estimator to Gaussian and paid the required extra number of rounds to ensure that the approximation error does not impact the overall estimation accuracy.  
      \item Considering the overall package of the estimation accuracy and the number of slots required for estimation, GEAN performs better than previously proposed estimation schemes.  
      \item We have presented the analysis of GEAN under two different channel models i.e. $\{0, 1\} $ and  $\{0, 1,e  \} $ along with their respective performances.  
   \end{enumerate}
   
   It should be noted that a conference version of this work \cite{hasan2018estimation} has been  published in the proceedings of IEEE International Conference on Communications (IEEE ICC 2018). The major difference between the two versions is that the conferecne version presents a brief analysis and limited results only on $ \{0,1\} $ channel model. It does not cover $ \{0,1,e\} $ channel model and the detailed derivations and discussions on the $ \{0,1\} $ channel model that this journal version has. The next paragraph gives a brief outline about the organization of the rest of the paper. 

Since the analysis of GEAN both under  $ \{0,1\} $ and  $ \{0,1,e\} $ channel models follow similar paths, sections II, III and IV combine together to set the rigorous analytical ground work of the paper under $ \{0,1\} $ channel model, which we use as the framework when we extend our analysis to $ \{0,1,e\} $ channel model in section V. Section VI talks about the algorithm the paper used to get an upper bound on the tag population size. Comparative  performance of GEAN with some of the recently proposed schemes and conclusion are given in Sections VII and VIII respectively. At the end we have references and appendices containing all the proofs of the lemmas and theorems used in the paper. 

\section{Formulation of the problem}
Before we jump into the mathematical formulation of the problem, it would serve well to introduce the protocol we used and give a brief outline of our approach. Expectedly a good number of works have  proposed different techniques to solve the mentioned estimation problems. As a result there already have been many communication protocols in literature. Though the idea of our work is independent of any such protocol, to assess the worth of our method we have resorted to the protocol explained in the next paragraph which is one of  the widely used protocol in literature \cite{kodialam2007anonymous} \cite{han2010counting} \cite{shahzad2015fast}. 

We used the framed slotted Aloha protocol specified in C1G2 [standardized in EPCGlobal C1G2 RFID standard \cite{epcglobal2004radio} and implemented
in commercial RFID systems] as the MAC-layer communication protocol. To begin the process AP broadcasts the frame size $(f)$ and a random seed number $(S)$ to all the AGs in its vicinity. Each of the AGs participate in the forthcoming frame with probability $ p $, where $ p $ is the \textit{persistence probability}, the probability that decides if an AG is going to remain active to participate in the forthcoming frame. Each individual AG has an $ ID$ and uses $f$ and $S$ values to evaluate a hash function $h(f,S,ID)$. The value of the hash function is uniformly distributed in the range $[1,f]$. Each AG has a counter that has an initial value equal to the slot number that it has got evaluating the hash function. After each slot the AP sends out a termination signal to all the AGs and each AG decreases its counter value by 1. At any given point the AGs with counter value equal to 1, reply to the reader. The next paragraph gives a brief outline of our approach. 

At the beginning of the estimation process GEAN runs a probe using the Flazolet Martin algorithm \cite{flajolet1985probabilistic} to get a rough upper bound $ t_{m} $ on the AG population size $ t $. The critical parameters $ p $, $ f $ and $ n $ are calculated based on the upper bound $t_{m} $ and the accuracy requirements $ \alpha $, $ \beta $, where $ n $ is the number of rounds (i.e. frames) required to meet the accuracy requirements . Using standardized framed slotted Aloha protocol, GEAN gets a reader sequence of either  $\{0, 1 \} $ or $\{0, 1,e \} $ depending on the channel model. GEAN uses $\frac{N_{n}-N_{0}}{f}$ and $\frac{N_{e}-N_{1}}{f}$  as the estimators  under $\{0, 1 \} $ and $\{0, 1,e \} $ channel models respectively. Where  $N_{n}$ is the number of non-empty slots,  $N_{0}$ is the number of $ 0 $s, $N_{1}$ is the number of slots having exactly one reply and $N_{e}$ is the number of slots having more than one replies in the reader sequence.  GEAN calculates the value of the  respective estimator  for each round and after $n$ rounds of these measurements  takes the average of all these values. This average is finally substituted for the true mean in the expected value equation of the respective  estimator to estimate the AG population size by an inverse function.  We have shown in this paper that the expected value function of GEAN under both channel models are invertible and analyzed the conditions under which  $ \frac{N_{n}-N_{0}}{f}$ and $ \frac{N_{e}-N_{1}}{f}$ are  asymptotically Gaussian distributed, while meeting the imposed estimation accuracy requirements. The required number of slots for estimation is defined as $(f +l)\times n$, where $l=1$ms (i.e. $\approx 3.33$ time slots) is the C1G2 specified mandatory time delay between the end of a frame and the start of the next one \cite{epcglobal2004radio}, \cite{reza2011rfid}. In the next two paragraphs, we will explain two important underlying  assumptions of the protocol we used. 

Firstly, in  our above mentioned framework AP identifies the responses coming from the AGs and generate the sequence of $ \{0,1\} $ or $ \{0,1,e\} $. Asking \textquotedblleft to what extent can we trust that generated sequence?\textquotedblright  would be a very legitimate question given the context. Because there can be numerous ways that can  result in a \textit{false positive} (a $ 0 $ getting detected as a $ 1 $) or a \textit{false negative} (a $ 1 $ getting detected as a $ 0 $). For example, the range of the antenna of the detection device might be extended by metallic objects and result in a false positive, and similarly an erroneous communication link due to environmental or physical interferences may result in a false negative. There have been good number of works in literature addressing either type of miss detection \cite{alfian2019false} \cite{massawe2012reducing}, and this part of the work is beyond the scope of our work. Essentially we are assuming the ideal scenario where each bit generated by the protocol can be trusted with probability $ 1 $, and such an assumption has been implicitly made by a predominant number of works in this trail \cite{kodialam2006fast}\cite{kodialam2007anonymous}  \cite{han2010counting} \cite{shahzad2015fast}. 

Secondly, in the original  protocol each AG, given  a frame size, selects one out of  $ f $ slots to respond. As a result, we can think of each  particular ID as producing a binary vector of dimension $f$, with one and only one of the $f$ entries set as $ 1 $ and the rest set as $ 0 $. The final outcome due to the aggregated behaviors of all IDs is the binary sum of these $ f $-dimensional binary vectors.  Apparently, the analysis of this model is highly complicated. However, a much more analytically tractable model where instead of picking  one out of $ f $ slots to reply an AG independently decides to reply in each slot of the frame regardless of its decision about previous or forthcoming slots,  has been proven to produce the same statistical outcome as the original model \cite{vogt2002efficient}. Since then it has been a norm to use this statistically equivalent model  \cite{cha2005novel}\cite{han2010counting}\cite{shahzad2015fast}. This slot by slot independence implies that an AG may end up choosing none or more than one slots in a frame. But the expected value of the number of slots that an AG chooses to reply in a frame is still $ 1 $ \cite{shahzad2015fast}. Since our scheme works with multiple frames of bigger frame sizes than the other schemes, we can asymptotically expect to observe this expected number. Hence the analysis remains the same with or without the assumption of independence. 

Having described our approach and its underlying assumptions, we are now well set to embark on the mathematical formulation of our work. Consistent with the above definitions and notations, equations \eqref{1} and \eqref{2} define  GEAN estimator and its expected value equation respectively, 
\begin{align}
&Z_{f}(t) \triangleq \frac{N_{n}-N_{0}}{f}\label{1}\\
&g_{f}(t) \triangleq E\left[\frac{N_{n}-N_{0}}{f}\right]=E\left[Z_{f}(t)\right]\label{2}
\end{align}

\begin{figure}
\centering
\includegraphics[scale=0.7]{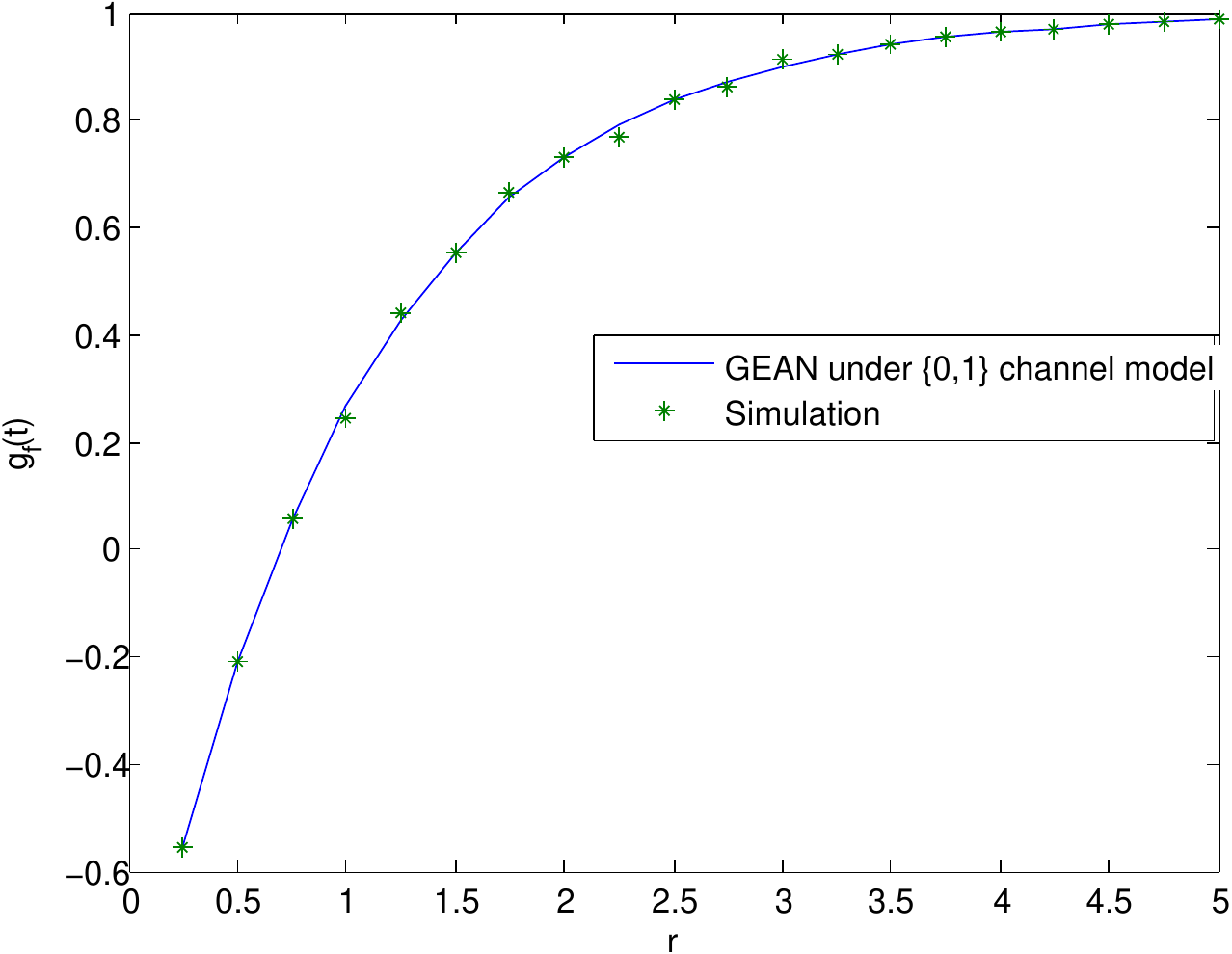}
\caption{Expected value function of GEAN $g_{f}(t)$ against   $r=\frac{tp}{f•}$ under $ \{ 0,1\}  $ channel model . $(f=200, p=1)$}
\label{e}
\end{figure}

\begin{figure}
\centering
\includegraphics[scale=0.7]{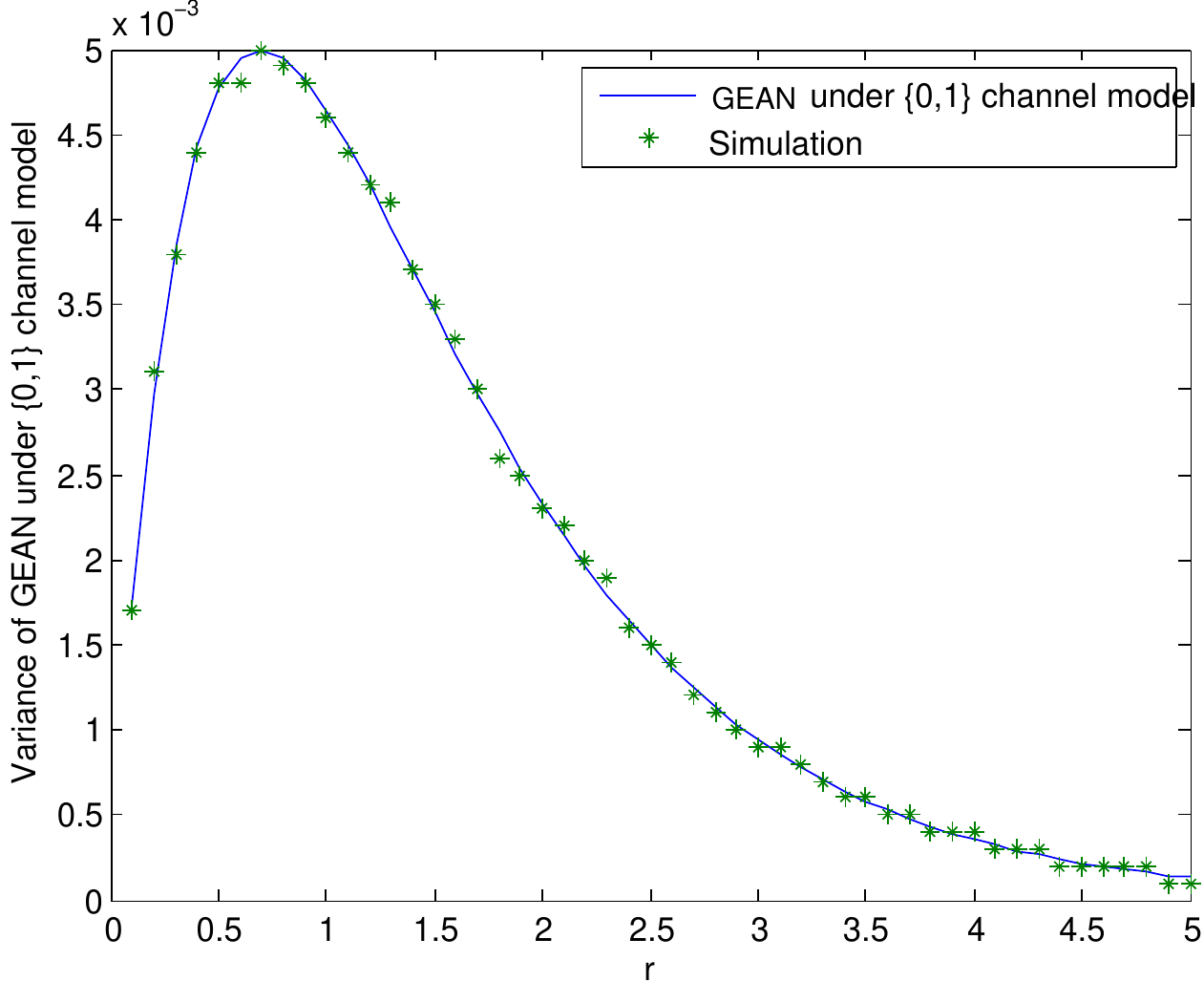}
\caption{Variance of GEAN estimator  against   $r=\frac{tp}{f•}$ under $\{0,1\}$ channel model. $(f=200, p=1)$}
\label{v}
\end{figure}
 
 As we already mentioned in the preceeding paragraph that, to make the formal development of our work tractable we assumed independence of slots. Let $X_{ij}\sim$ Bernoulli$(\frac{p}{f})$ be the random variable that represents the probability that the $i$th AG replies to the $ j^{th}$ slot. So the value of $X_{ij}$ is $1$ with a probability $\frac{p}{f}$ and the value of $X_{ij}$ is $0$ with a probability $(1-\frac{p}{f})$, i.e,

\begin{align}
X_{ij}=
    \begin{cases}
       1, &\text{with probability}\ \frac{p}{f}  \\
       0, &\text{with probability}\ (1-\frac{p}{f})  \\
    \end{cases}
\end{align}

Now we introduce $Y_{j}$  to represent the random variable wheather $j$th slot of the frame is empty or non-empty. i.e. we define,

\begin{align}\label{4}
Y_{j}=\sum_{i=1}^{t}X_{ij} .
\end{align} 

It is straightforward to derive the following probability distribution of $Y_{j}$,
\begin{align}\label{5}
p_{Y_{j}}(y)=
    \begin{cases}
      \left(1-\frac{p}{f}\right)^{t}=p_{0}, &  y=0  \\
    1-\left(1-\frac{p}{f}\right)^{t}=p_{n}, & y\neq 0\\
         \end{cases}
  \end{align}
\noindent  where $ p_{0}$ and $p_{n}$  are the probabilities that a slot empty and non-empty respectively. 
  
Introduction of the following two indicators $Y_{j}^{(n)}$ and $Y_{j}^{(0)}$ makes our analysis easier, 
\begin{align}\label{y}
Y_{j}^{(n)}=
    \begin{cases}
       1, &\text{when}\ Y_{j} \neq 0  \\
       0, &\text{when}\ Y_{j}=0 
    \end{cases} ,\quad
Y_{j}^{(0)}=
    \begin{cases}
       1, &\text{when}\ Y_{j}=0  \\
       0, &\text{when}\ Y_{j}\neq 0 \\
    \end{cases}
  \end{align}
So, \eqref{1} can be re-written as
\begin{align}\label{8}
\Rightarrow Z_{f}  = \frac{1}{f}\sum_{j=1}^f  Z_{j,f}
\end{align}\
where
\begin{align}\label{9}
 Z_{j,f} \triangleq Y_{j}^{(n)} - Y_{j}^{(0)}.
\end{align}
It is straightforward to show that $Z_{j,f}$ has the following PMF. 
\begin{equation}\label{a1}
    Z_{j,f}=
    \begin{cases}
    1, & \text{with probability}\ p_{n}\\
     -1, & \text{with probability}\ p_{0}\\
    \end{cases}
  \end{equation}
The first and second moments of $Z_{j,f}$ can be given by, 
\begin{align}
E[Z_{j,f}]&=\mu_{j,f}=p_{n}-p_{0},\label{12}\\
E[Z_{j,f}^{2}]&=p_{n}+p_{0}.\label{13}
\end{align}
Combination of \eqref{12} and \eqref{13} gives us the variance of $Z_{j,f}$,
\begin{align}\label{14}
\sigma_{j,f}^{2}=p_{n}+p_{0}-(p_{n}-p_{0})^{2}=1-(p_{n}-p_{0})^{2}.
\end{align}
Let, $\mu_{f}$ and $\sigma_{f}^{2}$ be the mean and variance of $Z_{f}$, respectively. Using \eqref{2}, \eqref{8}, \eqref{12} and \eqref{14} we have, 
\begin{align}
\mu_{f}&=g_{f}(t)=p_{n}-p_{0},\label{15}\\
\sigma_{f}^{2}&=\frac{1}{f}[p_{n}+p_{0}-(p_{n}-p_{0})^{2}].\label{16}
\end{align}
Now using the expressions from  \eqref{5},  equation \eqref{15} can be rewritten as,
\begin{align}\label{17}
g_{f}(t) &=   1- 2\left(1-\frac{p}{f}\right)^{t}
\end{align}
We define, 
\begin{align}\label{r}
r\triangleq\frac{tp}{f}
\end{align}
which essentially is the average number of AGs per slot.
\begin{lem}\label{l1}
  $ g_{f}(t) $ given in \eqref{17} is  a monotonically increasing function of $ r $ .
\end{lem}
The proof to Lemma \ref{l1} is given in Appendix A. 

\section{Gaussian Approximation of the Estimator}
Our estimation of the AG population size has to maintain the accuracy requirement given by the condition,
\begin{align}\label{ac}
P[|\hat{t}-t|\leq  \beta t] \geq \alpha
\end{align}
Since we are using $Z_{f}$ as our estimator to determine the value of $\hat{t}$, using \eqref{1} and \eqref{2}, the condition in \eqref{ac} can be written as, 
\begin{align}\label{ac1}
&P[|g_{f}^{-1}\{Z_{f}\}-t|\leq  \beta t] \geq \alpha\notag\\
\iff  &P[(1-\beta)t \leq g_{f}^{-1}\{Z_{f}\} \leq (1+\beta)t]\geq \alpha\notag\\
\iff  &P[g_{f}\{(1-\beta)t\} \leq Z_{f}  \leq g_{f}\{(1+\beta)t\}]\geq \alpha
\end{align}
Now to perform our estimation of the AG population size while maintaining the accuracy requirements given in \eqref{ac1}, we need the following two things, 
\begin{enumerate}
\item $g_{f}(t)$ has to be an invertible function.
\item a well approximated PDF for $Z_{f}(t)$. 
\end{enumerate}
In the previous section we saw that  $g_{f}(t)$  is monotonic function and hence invertible. This section is particularly devoted to the analysis of the conditions under which $Z_{f}(t)$ can be well approximated by a Gaussian random variable within a frame. Since the probabilities given in \eqref{5} vary with the frame size, we require 
a special version of the CLT, that is Triangular Array CLT \cite{araujo1980central}, to prove that $Z_{f}(t)$ follows Gaussian distribution. In other words, we  resort to Lindeberg Feller Theorem \cite{brown1971martingale}. \

The statement of Lindeberg Feller Theorem says, let $\{X_{n,i}\}$ be an  array of independent random variables with $E[X_{n,i}]=0$ and $E[X_{n,i}^{2}]=\sigma_{n,i}^{2}$, $Z_{n}=\sum_{i=1}^{n}X_{n,i}$ and $B_{n}^{2}=Var(Z_{n})=\sum_{i=1}^{n} \sigma_{n,i}^{2}$, then $Z_{n}\rightarrow N(0, B_{n}^{2})$ distribution if the condition below holds for every $\epsilon >0$,
\begin{equation}\label{lf}
\frac{1}{B_{n}^{2}} \sum_{i=1}^{n} E\left[  X_{n,i}^{2} 1_{X_{n,i}}\{ | X_{n,i}| > \epsilon B_{n}\}\right]  \rightarrow 0
\end{equation}
where $  1_{X}\{A\} $ is the indicator function of a subset A of the set X, and  is defined as, 
\begin{equation*}
    1_{X}\{A\}:=
    \begin{cases}
      1, &  x\in A\\
0 & x\notin A\\
    \end{cases}
  \end{equation*}

In our algorithm, it is easy to see that the set of random variables $\{Z_{j,f}\}$ given in  \eqref{a1} are independent.  From equations \eqref{12} and \eqref{15}  we see that $E[Z_{j,f}]=\mu_{f}$. Lindeberg Feller theorem requires the variable to have zero mean which is not the case with $Z_{j,f}$. To fulfill that requirement, we define a new set of  variables $\{\widetilde{Z}_{j,f}\}$ such that,
\begin{align}\label{if}
 \widetilde{Z}_{j,f}= Z_{j,f}-\mu_{f}
\end{align}
Now using \eqref{a1}, the probability distribution of  $\widetilde{Z}_{j,f}$ can be given by,
\begin{equation}\label{mf}
    \widetilde{Z}_{j,f}=
    \begin{cases}
1-\mu_{f}, & \text{with probability}\ p_{n}\\
-1-\mu_{f}, & \text{with probability}\ p_{0}\\
    \end{cases}
  \end{equation}
Using \eqref{12},\eqref{14} and \eqref{if} we have, 
\begin{align}
&E[\widetilde{Z}_{j,f}]=0\\
&Var[\widetilde{Z}_{j,f}]=\sigma_{j,f}^{2}
\end{align}
We define, 

\begin{align}\label{sf}
S_{f} & \triangleq \sum_{j=1}^{f}\widetilde{Z}_{j,f}=\sum_{j=1}^{f}(Z_{j,f}-\mu_{f})
\end{align}

Now, $\{\widetilde{Z}_{j,f}\}$ are independent random variables with $E[\widetilde{Z}_{j,f}]=0$ , $S_{f}=\sum_{j=1}^{f}\widetilde{Z}_{j,f}$ and $Var[S_{f}]=\sum_{j=1}^{f}Var(\widetilde{Z}_{j,f})=f\sigma_{j,f}^{2}$. According to Lindeberg Feller Theorem, $S_{f}$ will be asymptotically $N(0,f\sigma_{j,f}^{2})$ if the condition below holds for every $\epsilon >0$,

\begin{equation}\label{lfc}
\frac{1}{f\sigma_{j,f}^{2}} \sum_{j=1}^{f} E\left[  \widetilde{Z}_{j,f}^{2} 1_{\{\widetilde{Z}_{j,f}\}}\{ | \widetilde{Z}_{j,f}| > \epsilon \sqrt{f}\sigma_{j,f}\}\right]  \rightarrow 0
\end{equation}
Substituting $S_{f} \sim N(0,f\sigma_{j,f}^{2})$ in \eqref{sf}, simple algebraic manipulations using \eqref{8}, \eqref{14}, \eqref{16} and \eqref{sf} give us $Z_{f} \rightarrow N(\mu_{f},\sigma_{f}^{2})$. So, we can sum up, $Z_{f} \rightarrow N(\mu_{f},\sigma_{f}^{2})$ if \eqref{lfc} holds.

\begin{figure}
\centering
\includegraphics[scale=0.7]{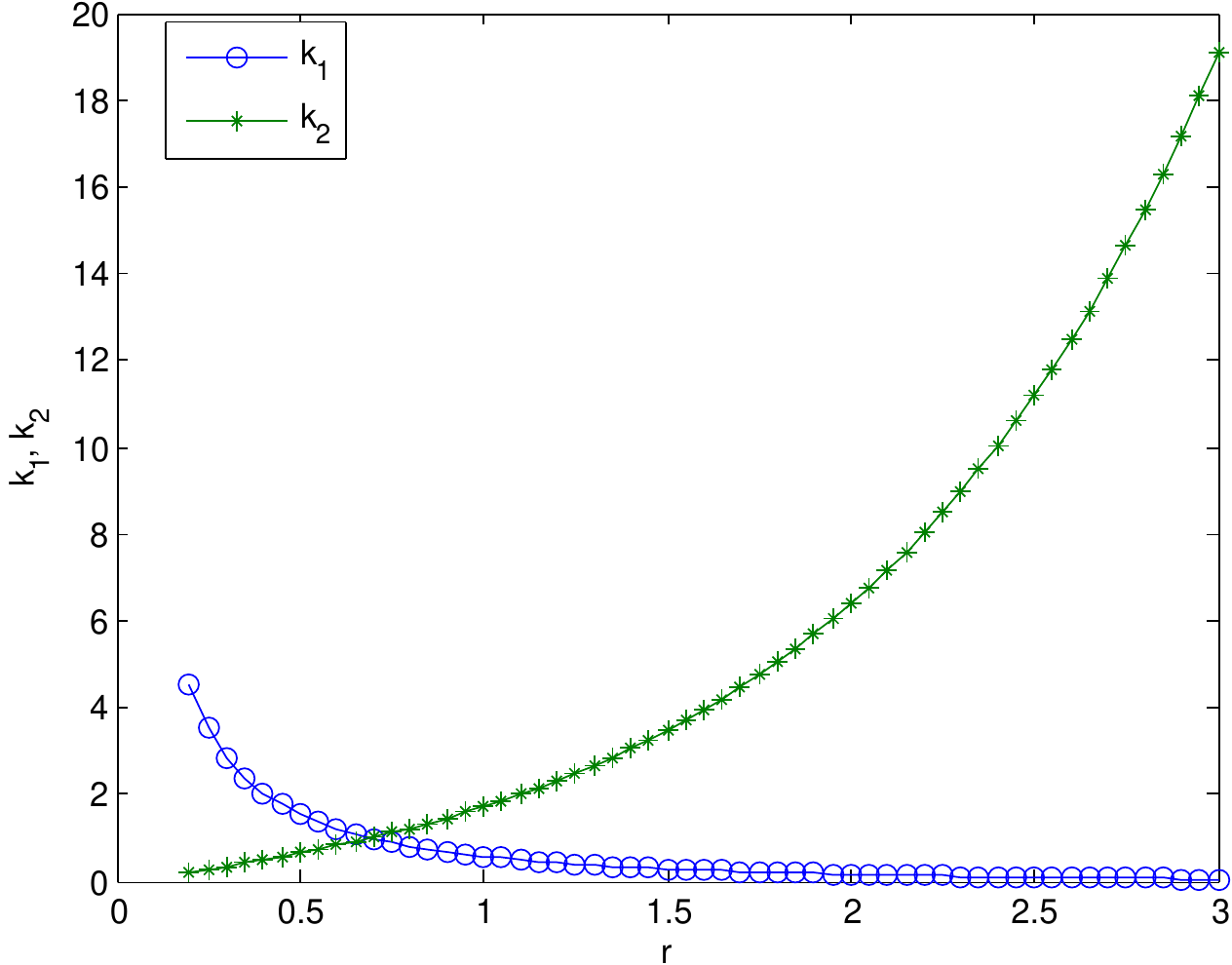}
\caption{$k_{1}, k_{2}$  against $r$, under $ \{0,1\} $ channel model}
\label{k01}
\end{figure}

In the condition given in \eqref{lfc} , the indicator function $1_{\{\widetilde{Z}_{j,f}\}}\{ | \widetilde{Z}_{j,f}| > \epsilon \sqrt{f}\sigma_{j,f}\}$ plays a pivotal role. For the variable $\widetilde{Z}_{j,f}$ we have the following $2$ cases of the indicator function, 
\begin{align}
& |1-\mu_{f}| > \epsilon \sqrt{f}\sigma_{j,f}\label{a2}\\
& | -1-\mu_{f}| > \epsilon\sqrt{f}\sigma_{j,f}\label{a3}
\end{align}
It is easy to see that if none of \eqref{a2} and \eqref{a3}  holds,  \eqref{lfc} not just converges to but actually becomes $ 0 $. We have shown in  Appendix B that if the following condition holds, none of the  \eqref{a2} and \eqref{a3} holds, or consequently \eqref{lfc} holds,
\begin{align}\label{33}
 \epsilon^{2}f \geq k(r)
\end{align}
where $k(r)$ is defined as 
\begin{align}\label{kaka}
 k(r)\triangleq max\{k_{1}(r), k_{2}(r)\}
\end{align}

and the values for $k_{1}$ and $ k_{2}$ are given by the following equations:

\begin{align}
&k_{1}(r) =\frac{e^{-r}}{1-e^{-r}}\\
&k_{2}(r) =\frac{1-e^{-r}}{e^{-r}}
\end{align}
Figure~\ref{k01} demonstrates $k_{1}$ and $ k_{2}$ against different values of $r$. We now come down to a condition, if \eqref{33} holds, \eqref{lfc} strictly becomes $ 0 $. So, for given $ r $ and $ \epsilon $  if we select a frame size large enough so that  \eqref{33} holds, the distribution of the estimator can be approximated as $Z_{f}\sim N(\mu_{f}, \sigma_{f}^{2})$.

\subsection{Quality Considerations of Gaussian Approximation}
The quality of the approximation depends on the value of the approximation error $ \epsilon $. To ensure that we satisfy the reliability requirements, we take  this approximation error into account when we calculate the overall estimation error. Exactly speaking $Z_{f}\rightarrow N(\mu_{f}, \sigma_{f}^{2})$ means, if the frame size is large enough to satisfy \eqref{33}, 

\begin{align}\label{q1}
&\left| P\left[ l\leq \frac{Z_{f}-\mu_{f}}{\sigma_{f}•}  \leq u \right] -P\left[ l\leq \theta \leq u \right]\right| \leq \epsilon
\end{align}
where, $\theta \sim N(0,1)$. Using the above equation we can write,

\begin{align}\label{q2}
 P\left[ l\leq \frac{Z_{f}-\mu_{f}}{\sigma_{f}•}  \leq u \right] \geq P\left[ l\leq \theta \leq u \right] - \epsilon
\end{align}
Now for the given reliability requirement $ \alpha $, using \eqref{q2} we have, 
\begin{align}\label{q3}
&P\left[ l\leq \theta \leq u \right] - \epsilon \geq \alpha\notag\\
&\Rightarrow P\left[ l\leq \theta \leq u \right]  \geq \alpha+\epsilon 
\end{align}
Which means that, if we approximate $ \frac{Z_{f}-\mu_{f}}{\sigma_{f}•}  $ as a standard normal, to compensate for the approximation error our target reliability should be $ \alpha+\epsilon $ instead of $ \alpha $. Using the fact that probability can not be greater than $ 1 $, 
\begin{align}
&\alpha+\epsilon\leq 1 \Rightarrow \epsilon_{max} = 1-\alpha\label{q4}
\end{align}

Equation \eqref{q4} gives the maximum value of approximation error that GEAN  can allow for a given reliability requirement $ \alpha $ . 
\subsection{Impact of the Approximation Error in Overall Estimation Accuracy} 
The $ \epsilon $ given above quantifies the the approximation error of the estimator for a given set of parameters i.e. $ (f, t, r)$. For example for given $\alpha=0.95, \beta=0.05, t=1200$ if we operate on $r=0.84 $ and select frame size $ f=200 $ the value of $ \epsilon $ calculated from \eqref{33} becomes $ 0.08 $. For the same set up if we select $f=1000  $ then we get $ \epsilon=0.03 $. In the first case we make more approximation error than in the second case. Equation\eqref{n} suggests that for $ f=200 $ we have to run more rounds than for $ f=1000 $. If we ignore the approximation error $ \epsilon $ and select $ f=200 $ for the above case, we are essentially aiming at a maximum achievable reliablility of $ 92\% $ as calculated from \eqref{q4}, where as our required reliability is $ 95\% $. But for $ f=1000 $ the maximum achievable reliablility is $ 97\% $ which means the required $ 95\% $ is achievable. So, its critical to exactly quantify the approximation error you make in terms of the estimation parameters of your algorithm to make sure you are aiming at a maximum reliability calculated from \eqref{q4} which is greater than your required reliability. In the performance section of this paper we clearly presented the significant hand that approximation error plays in the overall estimation accuracy. 
 \section{Selection of  Critical Parameters}
 \begin{figure}[!h]
 \removelatexerror
  \begin{algorithm}[H]\label{alg2}
   \caption{Estimate AG Population $(\alpha, \beta)$}
\textbf{Input:} 
\begin{enumerate}
\item Required reliability $\alpha$
\item Required confidence interval $\beta$
                          \end{enumerate}
   \textbf{Output:} Estimated AG population size $\hat{t}$ \\
Calculate $t_{m}$ := \textbf{upper bound}, and substitute $ t $ by $ t_{m} $.\\
Calculate $ r_{max} $ from \eqref{rmax} and discritize the range $ R=(0, r_{max}]  $\
calculate, $ l_{R}=length(R) $.

\For{$m:=1:l_{R}$}
{

Calculate $ k(r)  $ from \eqref{nk} and use that to obtain $f_{max}$ and $ f_{min} $ using \eqref{ffmax} and \eqref{ffmin} respectively for given $ r=R(m) $.\

make the array, $f_{array}=[f_{min}, f_{max}]$ \

calculate, $ l_{f}=length(f_{array}) $

\For{$i:=1:l_{f}$}
{
calculate $ p_{i} $ and $ \epsilon_{i} $ from \eqref{r} and \eqref{33} respectively for given $ r $ and $ f=f_{array}(i) $.\\
      Evaluate $n_{i}$.\\
}    
    Obtain $f_{m}$, and $n_{m}$ such that $(f_{m}+l)\times n_{m} := \min_{i} \{ (f_{array}(i)+l)\times n_{i}\}$\\
    
}   

 Obtain $ r_{op} $, $f_{op}$, and $n_{op}$ such that $(f_{op}+l)\times n_{op} := \min_{m} \{ (f_{m}+l)\times n_{m}\}$\\
calculate $ p_{op} $ and $ \epsilon_{op} $ from \eqref{r} and \eqref{33}\\

\For {$j:=1:n_{op}$}
{
Provide the reader with frame size $f_{op}$, persistence probability  $p_{op}$, and random seed $S_{j}$.\\
Run Aloha on the $ j $th frame.\\

Obtain $Z_{f}(j)= \frac{N_{n}-N_{0}}{ f_{op}}$ for the $ j $th frame\\
}
$\bar{Z_{f}} \leftarrow \frac{1}{n_{op}}\sum_{j}^{n_{op}} Z_{f}(j)$\\
Set $g_{f}(t):=\bar{Z_{f}}$ and solve \eqref{17} to get the estimated value $\hat{t}$ for AP population size $t$. \\
 \textbf{return} $\hat{t}$
\end{algorithm}
\end{figure}

\begin{figure}
\centering
\includegraphics[scale=0.7]{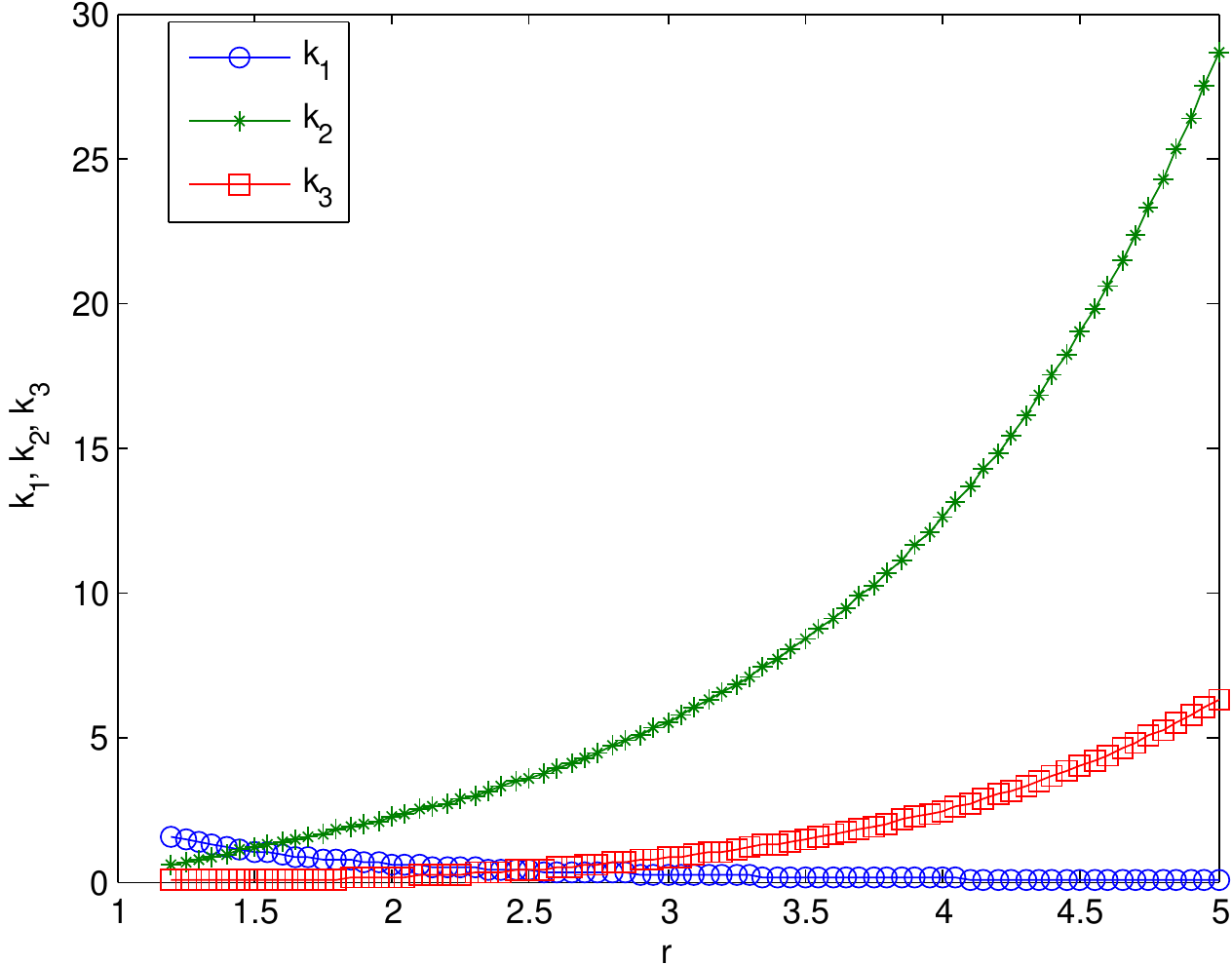}
\caption{$k_{1}, k_{2}, k_{3}$ against $r$ under \{0,1,e\} channel model}
\label{k}
\end{figure}

This section clarifies the steps to   attain the optimum parameters for GEAN under $ \{0,1\} $ channel model, described by Algorithm~\ref{alg2}. Since all the parameters are functions of the AG population size $ t $ which is unknown, we have substituted $ t_{m} $ for $ t $ in those equations. \

\subsection{Frame size $ f $}
For  given $ r $ and $ t_{m} $   we can have different pairs of  $ p $ and $ f $ satisfying \eqref{r}. Since $ p\leq 1 $, \eqref{r} should give us the maximum possible frame size when $ p=1 $ for  given $ r $ and $ t_{m} $. Using \eqref{r},
\begin{align}\label{ffmax}
f_{max}=\frac{t_{m}}{•r}
\end{align}
For a given $ r $ we will have a particular value of $ k(r) $ calculated from \eqref{kaka}, to satisfy the Lindeberg Feller conditions. Hence, we can get different pairs of $ f $ and $ \epsilon $ satisfying \eqref{33}. Plugging in maximum possible $ \epsilon $ given by \eqref{q4} in \eqref{33} should give us the minimum allowable frame size for given $ r $. Using \eqref{33}, 
\begin{align}\label{ffmin}
f_{min}=\frac{k(r)}{\epsilon_{max}^{2}•}
\end{align}
Because of the fact that we operate on a fixed value of $ r $, any frame size in the range $ [f_{min} , f_{max}]$ will have a corresponding $ \epsilon \leq \epsilon_{max}$, hence the pair will always satisfy Lindeberg Feller conditions.

Since $ f_{max} $ corresponds to $ p=1 $, any $ f $ in the range  $ [f_{min} , f_{max}]$ will have a corresponding $ p\leq 1 $ for given $ r $. 

\subsection{ Number of rounds $n$}
For a given frame size $f$, the required number of rounds required for the estimation of the tag population follows from the accuracy requirements specified in \eqref{ac1}. For all the frame sizes in the permissible range $[f_{min}, f_{max}]$, $g_{f}(t)$ is a monotonic function and $Z_{f}\sim N(\mu_{f}, \sigma_{f}^{2})$ with an approximation error $ \epsilon $.  So for any particular  $ f $ in that range, using \eqref{ac1} and \eqref{q2} we have, 
\begin{align}\label{ppp}
&P\left[\frac{g_{f}\{(1-\beta) t_{m}\} -\mu_{f}}{\sigma_{f}}\leq  \frac{Z_{f}-\mu_{f}}{\sigma_{f}} \leq \frac{g_{f}\{(1+\beta) t_{m}\}- \mu_{f}}{\sigma_{f}}\right]\notag\\ &  \geq \alpha+\epsilon
\end{align}
Since $Z_{f}\sim N(\mu_{f}, \sigma_{f}^{2})$, we have $\frac{Z_{f}-\mu_{f}}{\sigma_{f}}\sim N(0,1)$. Now taking the value of $\alpha+\epsilon$  as the CDF of standard normal distribution we get the corresponding cut off points in the standard normal curve  either side of the vertical axis. Let  the symmetric cutoff points to the right and to the left be  $z^{*}$ and $-z^{*}$ respectively. Using \eqref{ppp} it is straightforward to find the value of $ z^{*} $ to be $ z^{*}= Q^{-1}\left[\frac{1-\alpha -\epsilon}{2•} \right] $, where $ Q(z^{*})$ is the error fucntion. 
Let $n_{right}$ and $n_{left}$ be the number of rounds required corresponding to $z^{*}$ and $-z^{*}$ respectively. Because of the fact that the standard deviation gets scaled down $\sqrt{n}$ times if we take $n$ rounds of the measurements, solving the following two equations should give us the values for $n_{left}$ and $n_{right}$ , 
\begin{align}
\frac{g_{f}\{(1-\beta) t_{m}\} -\mu_{f}}{\frac{\sigma_{f}}{\sqrt{n_{left}}}}&=-z^{*}\label{53} \\
\frac{g_{f}\{(1+\beta) t_{m}\} -\mu_{f}}{\frac{\sigma_{f}}{\sqrt{n_{right}}}}&=z^{*}\label{54}
\end{align}
Since the two equations are not quite symmetric, the values of $n_{left}$ and $n_{right}$ might differ from each other. We will go by the higher value and take the ceiling if it is not an integer to ensure that we fulfill the minimum accuracy requirements. So, using \eqref{53} and \eqref{54} the required  number of rounds for given frame size $f$, is given by
\begin{align}\label{n}
n&=\lceil max\{n_{left}, n_{right}\} \rceil\notag\\
&= \left\lceil \frac{(z^{*})^{2} \sigma_{f}^{2}}{•\left[g_{f}\{(1-\beta) t_{m}\} -\mu_{f} \right]^{2}}, 
\frac{(z^{*})^{2} \sigma_{f}^{2}}{•\left[g_{f}\{(1-\beta) t_{m}\} -\mu_{f} \right]^{2}} \right\rceil
\end{align}
\subsection{Selection of $ r $}
All other parameters  were selected for a given value of $ r $. Calculation of the maximum possible $ r $ that follows from the fact that we need to ensure that the following holds,
\begin{align}\label{rcon}
f_{max} \geq f_{min}
\end{align}
Combinig \eqref{ffmin} and \eqref{rcon}, we can calculate the biggest $ r $ for which \eqref{rcon} holds, 

\begin{align}\label{rmax}
r_{max}= \sup_{r> 0} \{\epsilon_{max}^{2}f_{max}(r) \geq k(r)\}
\end{align}

So, for given $ t_{m} $ the range of the values of $ r $ that we operate on  is $(0, r_{max}] $. It is easy to see that, for any value of $ r_{max} $  the range $(0, r_{max}] $ exists. Since $ r$ can be arbitrarily small in that range, the maximum frame size $ f_{max} $ can be arbitrarily big and the condition $ f_{max}\geq f_{min} $ is always satisfied.  This  implies  we can estimate any arbitrary  AG population size under GEAN $ \{0,1\} $ channel model.

\section{GEAN under $ \{0,1,e\} $ channel model}
This section of the paper introduces $ \{0,1,e \} $ channel model and analyzes our scheme under this particular channel model. Though  the major part of the analysis under $ \{0,1,e\} $  will be presented with reference to that of $ \{0,1\} $ channel model, in terms of idea and outcome $ \{0,1,e\} $ is a substantial entity by itself. 

Under this model a $ 0 $ represents an empty slot, a $ 1 $ represents a  singleton slot and an $ e $ represents a collision slot. The Aloha protocol that we used for $ \{0,1\} $  is the same protocol that we use for $ \{0,1,e\} $ channel model to obtain the reader sequence. The estimator that we are using for $ \{0,1,e\} $ model is $\frac{•N_{e}-N_{1}}{•f}  $ where $ N_{e} $ is the number of collision slots and $ N_{1} $ is the number of singleton slots in a frame. The corresponding equations for \eqref{1} and \eqref{2} for this channel model can be given by, \eqref{n1} and \eqref{n2} respectively. 
\begin{figure}
\centering
\includegraphics[scale=0.7]{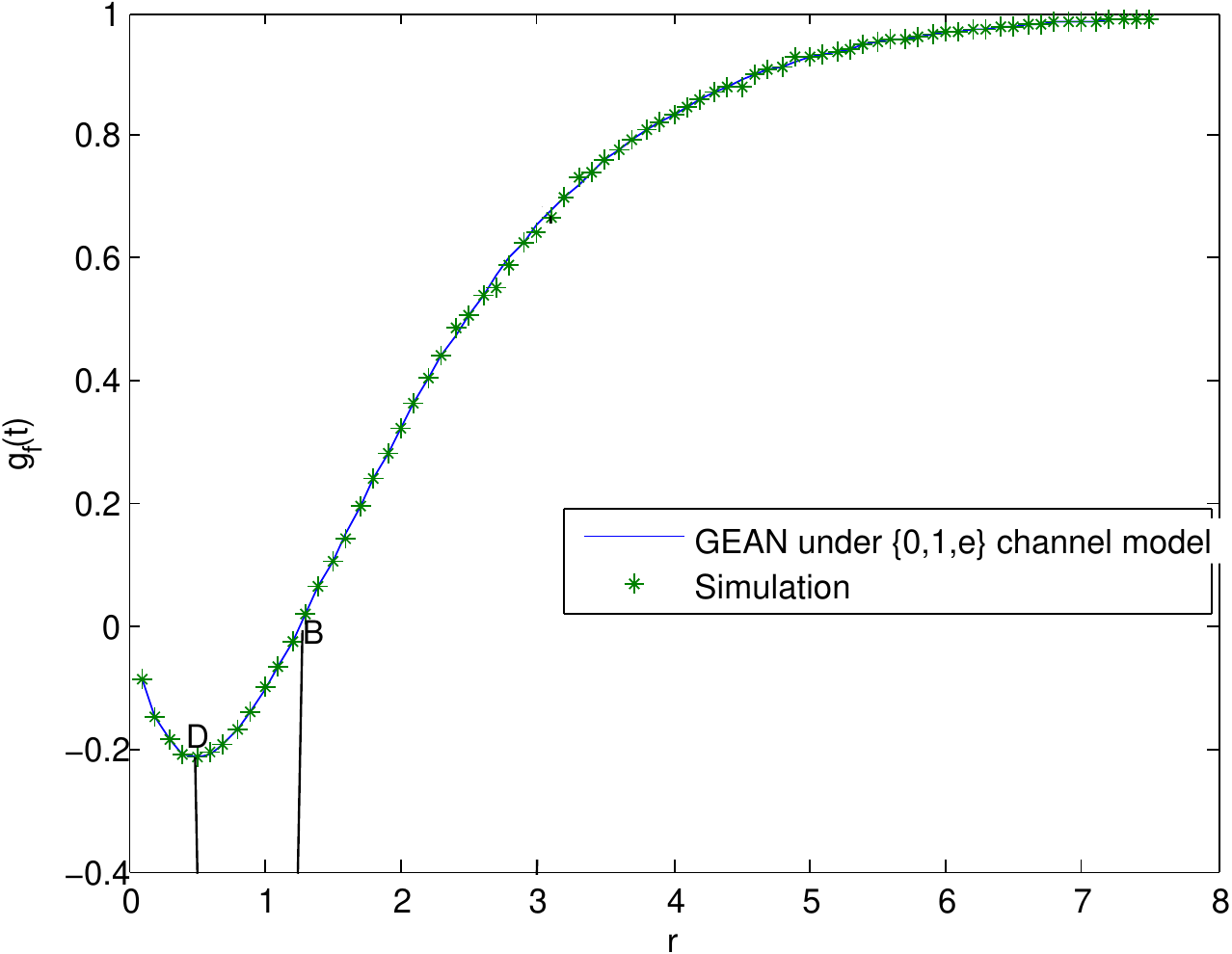}
\caption{Expected value function of GEAN $g_{f}(t)$ against   $r=\frac{tp}{f•}$ under $ \{ 0,1,e\}  $ channel model . $(f=200, p=1)$}
\label{e01}
\end{figure}
\begin{figure}
\centering
\includegraphics[scale=0.7]{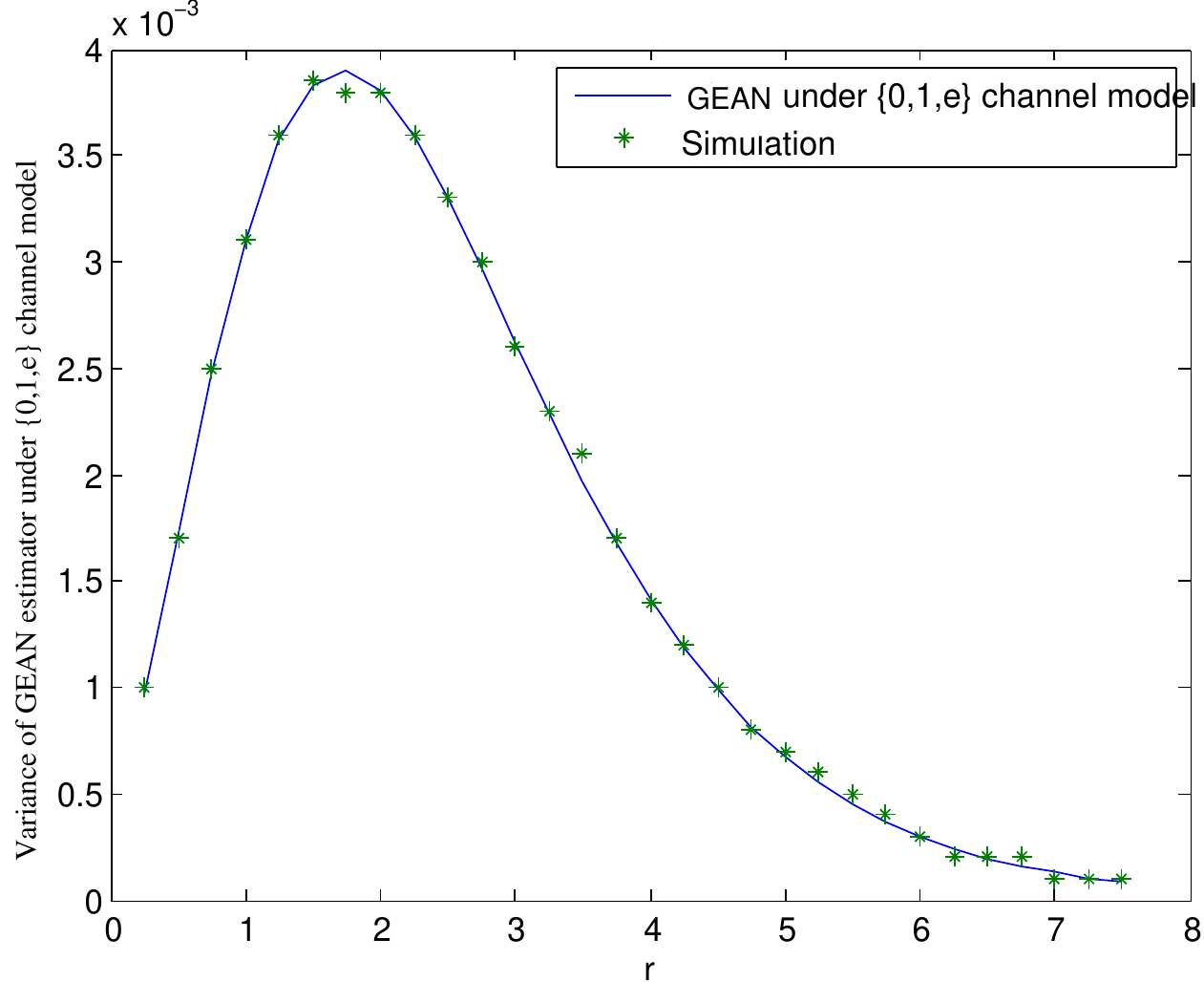}
\caption{Variance of GEAN estimator  against   $r=\frac{tp}{f•}$ under $\{0,1,e\}$ channel model. $(f=200, p=1)$}
\label{v01}
\end{figure}
\begin{align}
&Z_{f}(t) \triangleq \frac{N_{e}-N_{1}}{f}\label{n1}\\
&g_{f}(t) \triangleq E\left[\frac{N_{e}-N_{1}}{f}\right]\label{n2}
\end{align}

It is straightforward to see that the random variable $Y_{j}$ defined in \eqref{4}, has the following probability distribution under $ \{0,1,e\} $ channel model,
\begin{align}\label{n5}
p_{Y_{j}}(y)=
    \begin{cases}
      \left(1-\frac{p}{f}\right)^{t}=p_{0}, &  y=0  \\
    t\left(\frac{p}{f}\right)\left(1-\frac{p}{f}\right)^{t-1}=p_{1}, & y=1\\
     1-p_{0}-p_{1}= p_{e}, & y>1\\
    \end{cases}
  \end{align}
  where $ p_{0}, p_{1}$ and $ p_{e} $ are the probabilities that a particular slot is an empty, a singleton and a collision slot respectively. 
  Corresponding to the indicators defined  in \eqref{y}, we introduce the following two indicators for $ \{0,1,e\} $ channel model, 
\begin{align}
Y_{j}^{(e)}=
    \begin{cases}
       1, &\text{when}\ Y_{j}>1  \\
       0, &\text{when}\ Y_{j}\leq 1 
    \end{cases} ,\quad
Y_{j}^{(1)}=
    \begin{cases}
       1, &\text{when}\ Y_{j}=1  \\
       0, &\text{when}\ Y_{j}\neq 1 \\
    \end{cases}
  \end{align}
  Corresponding to \eqref{9}, we define $  Z_{j,f} \triangleq Y_{j}^{(e)} - Y_{j}^{(1)}$, and its easy to see that $  Z_{j,f}$ has the following probability distribution under $ \{0,1,e\} $ channel model,
  \begin{equation}\label{11}
    Z_{j,f}=
    \begin{cases}
      0, & \text{with probability}\ p_{0}\\
    -1, & \text{with probability}\ p_{1}\\
     1, & \text{with probability}\ p_{e}\\
    \end{cases}
  \end{equation}
  Derivation of the following mean and variance equations for $ Z_{j,f} $ under $ \{0,1,e\} $ channel model is straightforward. 
  \begin{align}
E[Z_{j,f}] &=   p_{e}-p_{1}\label{b1}\\
\sigma_{j,f}^{2}&=p_{e}+p_{1}-(p_{e}-p_{1})^{2}.\label{b2}
\end{align}
  It is important to notice that $\{0,1\} $ model equation \eqref{8} holds for $ \{0,1,e\} $ channel model as well. Using \eqref{8}, \eqref{n5}, \eqref{b1}, and \eqref{b2}, we get  the following mean and variance equations for $Z_{f}$ under $ \{0,1,e\} $ channel model. 
\begin{align}
g_{f}(t) &=   1- \left(1-\frac{p}{f}\right)^{t}- 2 t\left(\frac{p}{f}\right)\left(1-\frac{p}{f}\right)^{t-1}\label{n15}\\
\sigma_{f}^{2}&=\frac{1}{f}[p_{e}+p_{1}-(p_{e}-p_{1})^{2}].\label{n16}
\end{align}
 We notice in Figure~\ref{e01} that, there is a dip at the beginning and other than that the expectation curve is monotonically increasing.  The monotonically increasing part guarantees us a distinct inverse that will help us estimate the actual number of AGs. But for the dip we instead have a singularity i.e. we will get more than one horizontal points for one vertical point. We either have to operate in the monotonically increasing region, or we have to find a special way out to select the actual value out of the multiple values suggested by the estimator in the singularity region. The following two lemmas  shed more light on the matter. 
 Let the corresponding values of $t$ and $r$ at the point where the dip of $g_{f}(t)$ occurs be $ t_{LM} $ and $ r_{LM}$, respectively. \

\begin{lem}\label{lem1}
The local minimum of $g_{f}(t)$ curve occurs at an AG population size $ t_{LM} = \frac{f}{2p}$ or equivalently at $ r_{LM}=\frac{1}{2}$, given frame size $f$ and persistence probability $p$.
\end{lem}

\begin{lem}\label{lem2}
$g_{f}(t)$  is a convex function of $t$ for $t<\frac{3f}{2p}$, and for the rest of the $t$ values the function is concave, given frame size $f$ and persistence probability $p$.
\end{lem}
The proofs of Lemma~\ref{lem1} and Lemma~\ref{lem2} are given in  Appendix C and Appendix D respectively .\

Combining Lemma~\ref{lem1} and Lemma~\ref{lem2} we can see there exists  $\frac{1}{2p}f \leq t_{B} \leq \frac{3}{2p}f$ or equivalently $\frac{1}{2} \leq \frac{t_{B}p•}{f•} \leq \frac{3}{2}$ such that $g_{f}(t_{B})=g_{f}(1)$ (considering that the minimum possible AG population size is $ 1 $). That $ t $ corresponds to the point $ B $ in the $ g_{f} (t)$ curve and we represent  the corresponding value of $ r $ by $ r_{min} $. $g_{f}(t)$ will go into the singularity if we violate the following condition,  
 \begin{align}\label{mono}
r \geq r_{min}.
\end{align}
We numerically got the value of $r_{min}=1.2564$ . 
So, the summary from this section is that, if we have an $ r $ so that \eqref{mono} holds,  $g_{f}(t)$ will be a monotonic function and hence invertible. \\

It is important to note that, we do not know the value of $t$ rather we just have an upper-bound on $t$ which is $t_{m}$. The frame size we are selecting corresponds to $t_{m}$ not $t$.  Since our frame size corresponds to $t_{m}$ we may end up selecting a bigger $f$ than we should. As a result we may still end up operating in the singularity region of $ g_{f}(t) $. The next lemma helps us to resolve this issue. 
\begin{lem}\label{lem3}
In the case of a singularity, i.e. when we have to decide between two different $\hat{t}$ corresponding to the same value of the estimator, the value to the right of the dip is always the value desired. 
\end{lem}
The proof to Lemma~\ref{lem3} is given in the Appendix E .

 Now, corresponding to  \eqref{mf}, the modified version of $Z_{j,f}$, $\widetilde{Z}_{j,f}$ has the following probability distribution under $ \{0,1,e\} $ channel model,
\begin{equation}\label{nmf}
    \widetilde{Z}_{j,f}=
    \begin{cases}
      -\mu_{f}, & \text{with probability}\ p_{0}\\
-1-\mu_{f}, & \text{with probability}\ p_{1}\\
1-\mu_{f}, & \text{with probability}\ p_{e}\\
    \end{cases}
  \end{equation}
Using \eqref{if}, \eqref{b1} and \eqref{b2}  we have, 
\begin{align}
&E[\widetilde{Z}_{j,f}]=0\\
&Var[\widetilde{Z}_{j,f}]=\sigma_{j,f}^{2}
\end{align}
  It is obvious that under $ \{0,1,e\} $ channel model, the indicator function in \eqref{lfc} has  the following three cases, 
 \begin{align}
& |1-\mu_{f}| > \epsilon \sqrt{f}\sigma_{j,f}\label{30}\\
& | -1-\mu_{f}| > \epsilon\sqrt{f}\sigma_{j,f}\label{31}\\
&| -\mu_{f}| > \epsilon \sqrt{f}\sigma_{j,f}\label{32}
\end{align}
Consequently, the corresponding equation for $ k $ given in \eqref{kaka}, under $ \{0,1,e\} $ channel model would be, 
\begin{align}\label{nk}
 k(r)\triangleq max\{k_{1}(r), k_{2}(r), k_{3}(r)\}
\end{align}
and the values for $k_{1}, k_{2}$ and $ k_{3}$  are given by the following equations,
\begin{align}
&k_{1}(r) =\frac{1}{\left|\frac{-e^{r} \left( 1 +4r\right)}{\left(1+ 2r \right)^{2}}\right | -1}\\
&k_{2}(r) =\left|\frac { e^{2r} +  \left(1+2r \right) \left(  \frac{1}{4} - e^{r} \right)}{   \frac{1}{4}  \left(1+2 r\right)^{2}   - r e^{r}}\right| \\
&k_{3}(r) =\left|\frac{e^{2r} -2e^{r} \left(1+ 2 r \right) + \left(1+2r\right)^{2}}{\left(1+2r\right)^{2}  -e^{r} \left(1 +4 r \right)}\right| 
\end{align}
The proofs of the above  equations are given in Appendix F. 
Based on Lindeberg Feller theorem applied in $ \{0,1\} $ case, we can say, if we have a big enough frame size to satisfy \eqref{33} for given  $ \epsilon $ and a value of  $ k $  calculated from \eqref{nk}, \eqref{lfc} holds and we have $Z_{f}\sim$ asymptotically $N(\mu_{f}, \sigma_{f}^{2})$ under $ \{0,1,e\} $ channel model. To ensure that we take the approximation error into account, like $ \{0,1\} $ we have to maintain the target reliability $ \alpha+\epsilon $ instead of  $ \alpha $ for $ \{0,1,e\} $ channel model, and we still have that upper bound on the value of $ \epsilon $ given by \eqref{q4}.
\subsection{Critical parameters under $ \{0,1,e\} $ channel model}
We use the same equations \eqref{ffmax} and \eqref{ffmin} as under $ \{0,1\} $ channel model, to calculate $ f_{max} $ and $ f_{min} $ for given $ t_{m} $ and $ r $, except for the fact that the value of $ k(r) $ we use in \eqref{ffmin} is specific to $ \{0,1,e \} $ channel model and is calculated from \eqref{nk}. The corresponding values of $ p $ and $ \epsilon $ for each $ f $ in the range $ [f_{min}, f_{max} ]$ follow like    they did  under $ \{0,1\} $ channel model. \

Selection of the number of rounds  $ n $ required to meet the accuracy requirement under $ \{0,1,e\} $  is no different from that of $ \{0,1\} $ channel model. For any $ f $ in the $ f_{array} $ mentioned in Algorithm~\ref{alg:al}, we calculate  $ n $ required for the given accuracy requirements from \eqref{n},  as we did for the $ \{0,1\} $ channel model. 

 The only difference is that the fucntions $ g_{f}(t) $, $ \mu_{f} $ and $ \sigma_{f}^{2}$  we use in this case are the ones  specific to $ \{0,1,e \} $ channel model given by \eqref{n15} and \eqref{n16}.
 \begin{figure}[!h]
 \removelatexerror
  \begin{algorithm}[H]\label{alg:al}
   \caption{Estimate AG Population $(\alpha, \beta)$}
\textbf{Input:} 
\begin{enumerate}
\item Required reliability $\alpha$
\item Required confidence interval $\beta$
                          \end{enumerate}
   \textbf{Output:} Estimated AG population size $\hat{t}$ \\
Calculate $t_{m}$ := \textbf{upper bound}, and substitute $ t $ by $ t_{m} $.\\
Calculate $ r_{max} $ from \eqref{rmax01} and discritize the range $ R=[1.26, r_{max}]  $ \\
calculate, $ l_{R}=length(R) $.

\For{$m:=1:l_{R}$}
{

calculate $ k(r)  $ from \eqref{kaka} and use that to obtain $f_{max}$ and $ f_{min} $ using \eqref{ffmax} and \eqref{ffmin} respectively for given $ r=R(m) $.\

make the array, $f_{array}=[f_{min}, f_{max}]$ \

calculate, $ l_{f}=length(f_{array}) $

\For{$i:=1:l_{f}$}
{
calculate $ p_{i} $ and $ \epsilon_{i} $ from \eqref{r} and \eqref{33} respectively for given $ r $ and $ f=f_{array}(i) $.\\
      Evaluate $n_{i}$.\\
}    
    Obtain $f_{m}$, and $n_{m}$ such that $(f_{m}+l)\times n_{m} := \min_{i} \{ (f_{array}(i)+l)\times n_{i}\}$\\

}
Obtain $ r_{op} $, $f_{op}$, and $n_{op}$ such that $(f_{op}+l)\times n_{op} := \min_{m} \{ (f_{m}+l)\times n_{m}\}$\\
calculate $ p_{op} $ and $ \epsilon_{op} $ from \eqref{r} and \eqref{33}\\
\For {$j:=1:n_{op}$}
{
Provide the reader with frame size $f_{op}$, persistence probability  $p$, and random seed $S_{j}$.\\
Run Aloha on $ j $th frame.\\
Obtain $Z_{f}(j)= \frac{N_{e}-N_{1}}{ f_{op}}$ for the $ j $th frame\\
}
$\bar{Z_{f}} \leftarrow \frac{1}{n_{op}}\sum_{j}^{n_{op}} Z_{f}(j)$\\
Set $g_{f}(t):=\bar{Z_{f}}$ and solve \eqref{n15} to get the estimated value $\hat{t}$ for AG population size $t$. \\
 \textbf{return} $\hat{t}$
\end{algorithm}
\end{figure}

\subsection{Selection of $r$}
For $ \{0,1,e \} $ channel model, the value of $ r $ cannot be less than $r_{min}= 1.26 $ because of the singularity considerations.
 Hence, under $ \{0,1,e\} $ channel model the equation \eqref{rmax} for the maximum value of $ r $ gets modified to, 
\begin{align}\label{rmax01}
r_{max}= \sup_{r> r_{min}} \{\epsilon_{max}^{2}f_{max}(r) \geq k(r)\}
\end{align}
The value $ k(r) $ we use in   \eqref{rmax01} is the one specific to $ \{0,1,e\} $ channel model.\\

If for a given $ t_{m} $ the $ r_{max}$  is smaller than  $ r_{min} =1.26$, the   range $ [r_{min}, r_{max}] $ does not exist. That means the AG population size is not big enough to be estimated even at a value $ r=r_{min} $ for given $ \epsilon_{max} $. Plugging in $ r=r_{min} $ in \eqref{ffmax} and \eqref{ffmin} then solving the equations under the constraint $ f_{max}\geq f_{min} $, we have 
\begin{align}\label{tp}
t_{m}\geq \frac{k(r_{min})r_{min}}{\epsilon_{max}^{2}•}=t_{ml}
\end{align}
Equation~\eqref{tp} implies that, for a given accuracy requirement we can estimate an AG population size only if the upper bound is greater than $ t_{ml} $. For example, for $ \alpha=92\% $ i.e. $ \epsilon_{max}=0.08 $, we can only estimate  AG population sizes that have $ t_{m} $ greater than $ 230 $  under $ \{ 0,1,e\} $  channel model. 
    
\section{Upper bound on the AG population size $ t_{m}$}
Our algorithm requires an upper-bound on the AG population size which we obtain by using Flajolet and Martin's probabilistic counting
algorithm \cite{flajolet1985probabilistic}. We do it before calculating the parameters $p$, $f$, $n$ and $ r $. Because, to calculate all these parameters we need $t_{m}$. In Flajolet and Martin's probabilistic counting algorithm,  the AP keeps issuing one slot frames till it gets an empty slot. The persistence probability starts with a value $1$ and keeps on decreasing following a geometric distribution $(i.e., p=\frac{1}{2^{i-1}} \text{in the}\ i\text{th frame})$. If the empty slot occurs in the $j$th frame then, $t_{m}=1.2897 \times 2^{j-2}$ is considered to be an upper bound on the existing AG poputation size $t$ \cite{flajolet1985probabilistic}, \cite{qian2011cardinality}. Average of $t_{m}$ values obtained in large number of rounds assymptotically approaches $t$ \cite{flajolet1985probabilistic}. 

We want the upper bound to be as close to actual population size as possible. In \cite{shahzad2015fast} fair bit of analysis was done on how close $t_{m}$ is to $t$. Their result shows that for $99 \%$  reliability and $1 \%$ confidence interval $t_{m}$ is within $1.66 \times t$, and for $90\%$  reliability and $10 \%$ confidence interval , which is not particularly a very tight accuracy requirement, $t_{m}$ is within $1.83 \times t$. This gives us a fair idea that for reasonable accuracy requirements $t_{m}$ is within $2 \times t$. This is an assumption that we made in this paper that  $t_{m} \leq 2 \times t$ which is well supported by the findings in \cite{shahzad2015fast} and we take the average over $100$ rounds to get that $t_{m}$. 

%
%
%

\section{Performance Evaluation}
\begin{figure}
\includegraphics[width=\linewidth]{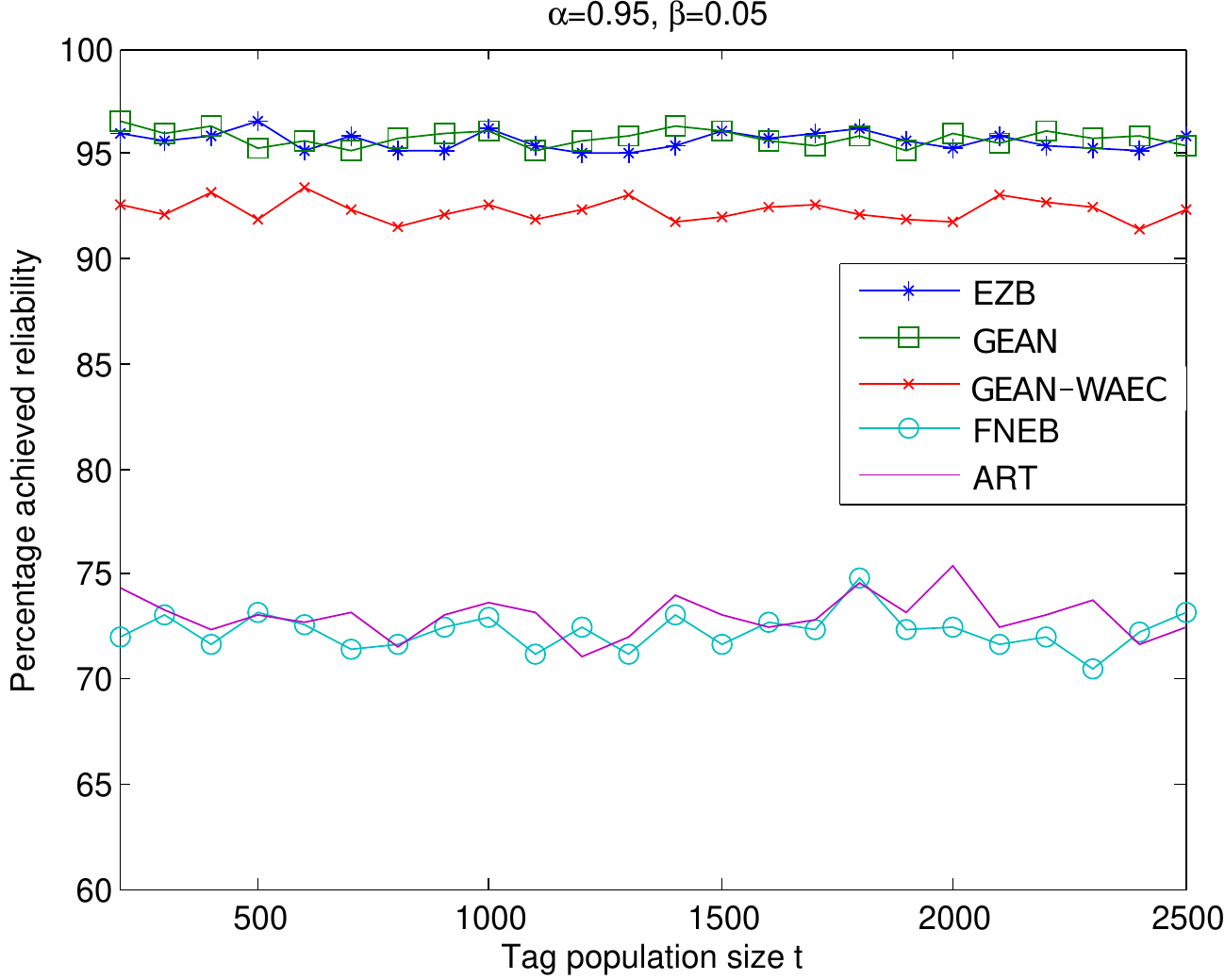}
\caption{Actual reliability achieved by different tag estimation schemes against different tag population sizes   $\alpha=0.95, \beta=0.05$}
\label{r01_f5}
\end{figure}

\begin{figure}
\includegraphics[width=\linewidth]{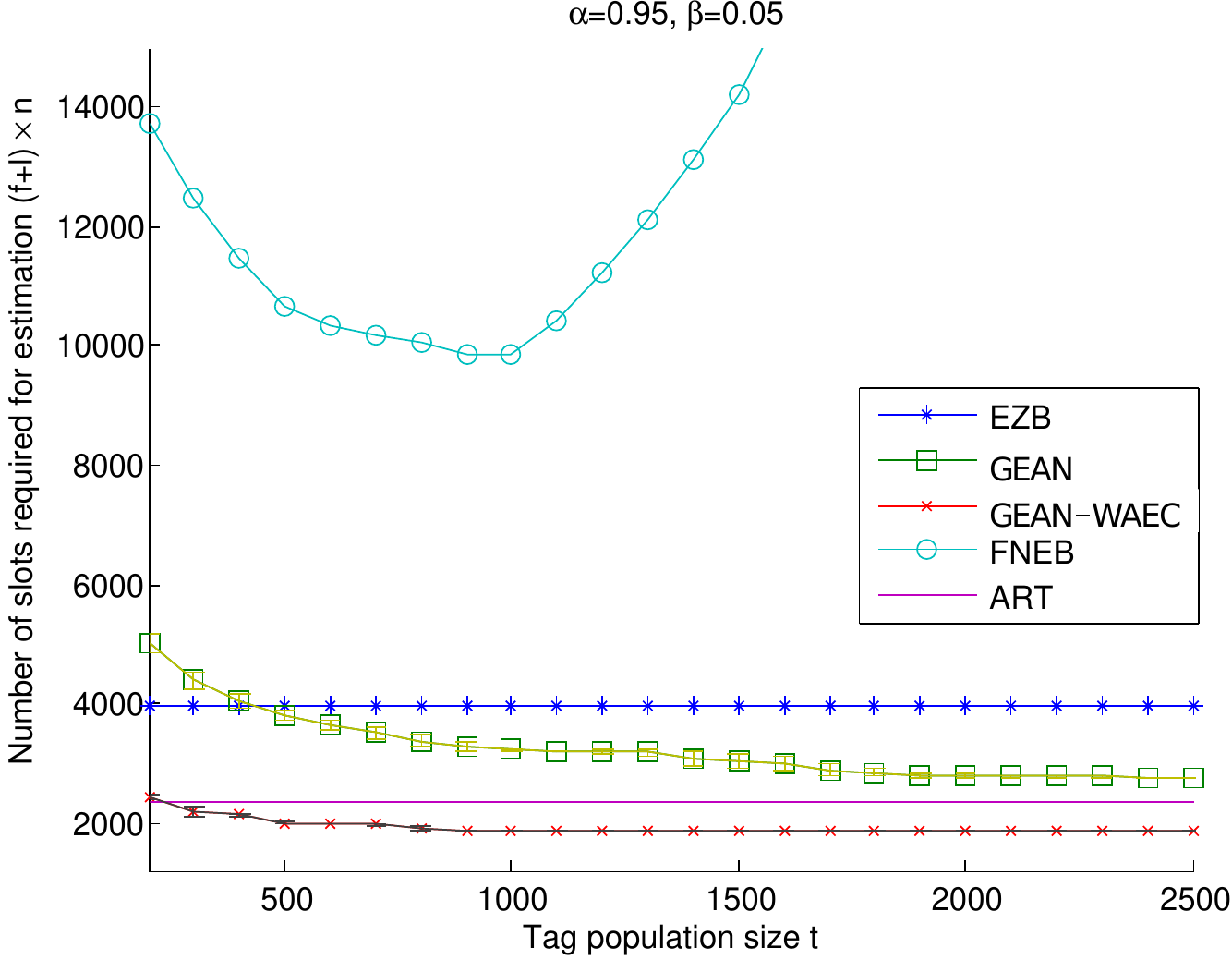}
\caption{Number of slots required by different tag estimation schemes against different tag population sizes  . $\alpha=0.95, \beta=0.05$}
\label{r01_f6}
\end{figure}

\begin{figure}
\includegraphics[width=\linewidth]{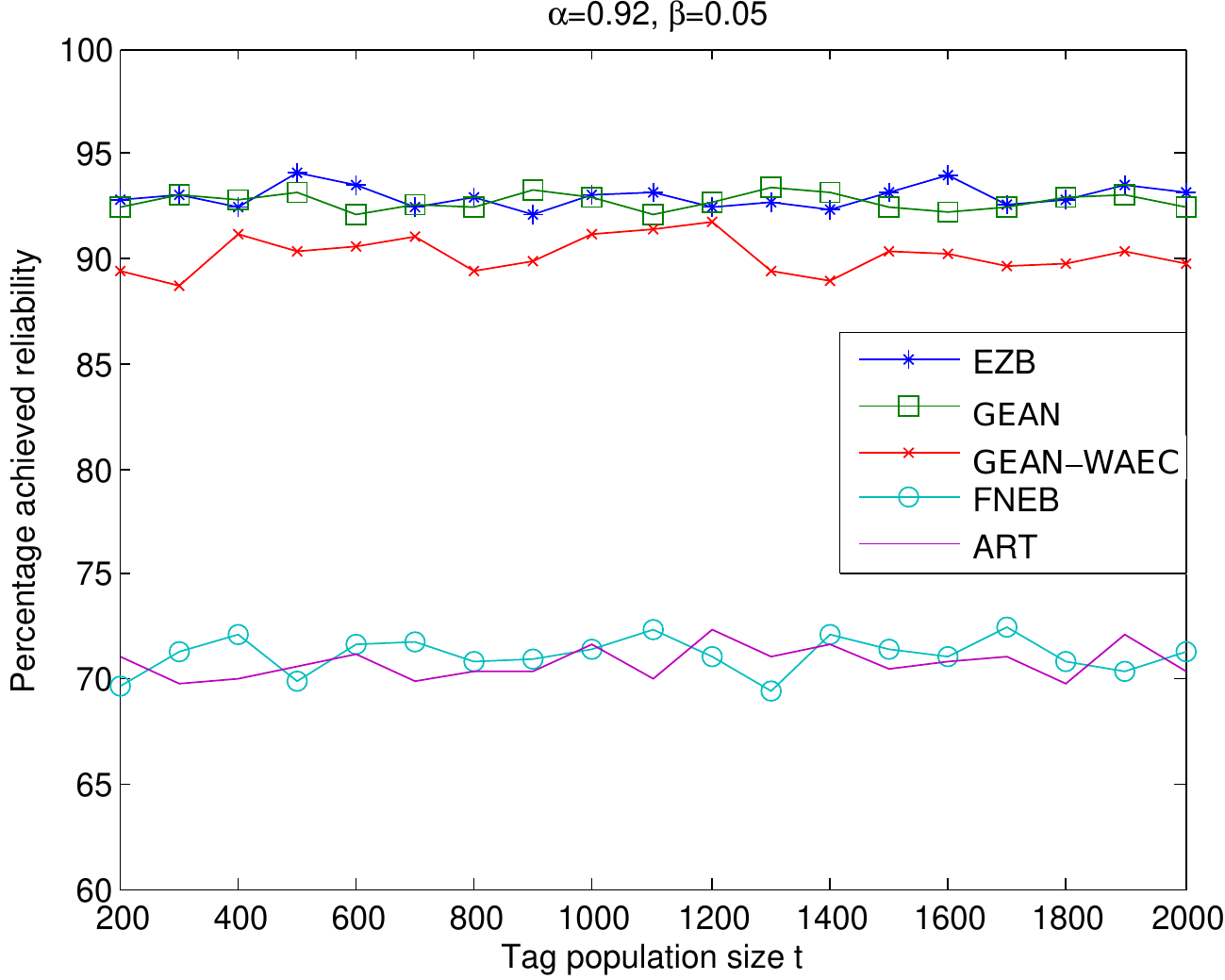}
\caption{Actual reliability achieved by different tag estimation schemes against different tag population sizes   $\alpha=0.92, \beta=0.05$}
\label{r01_f7}
\end{figure}

\begin{figure}
\includegraphics[width=\linewidth]{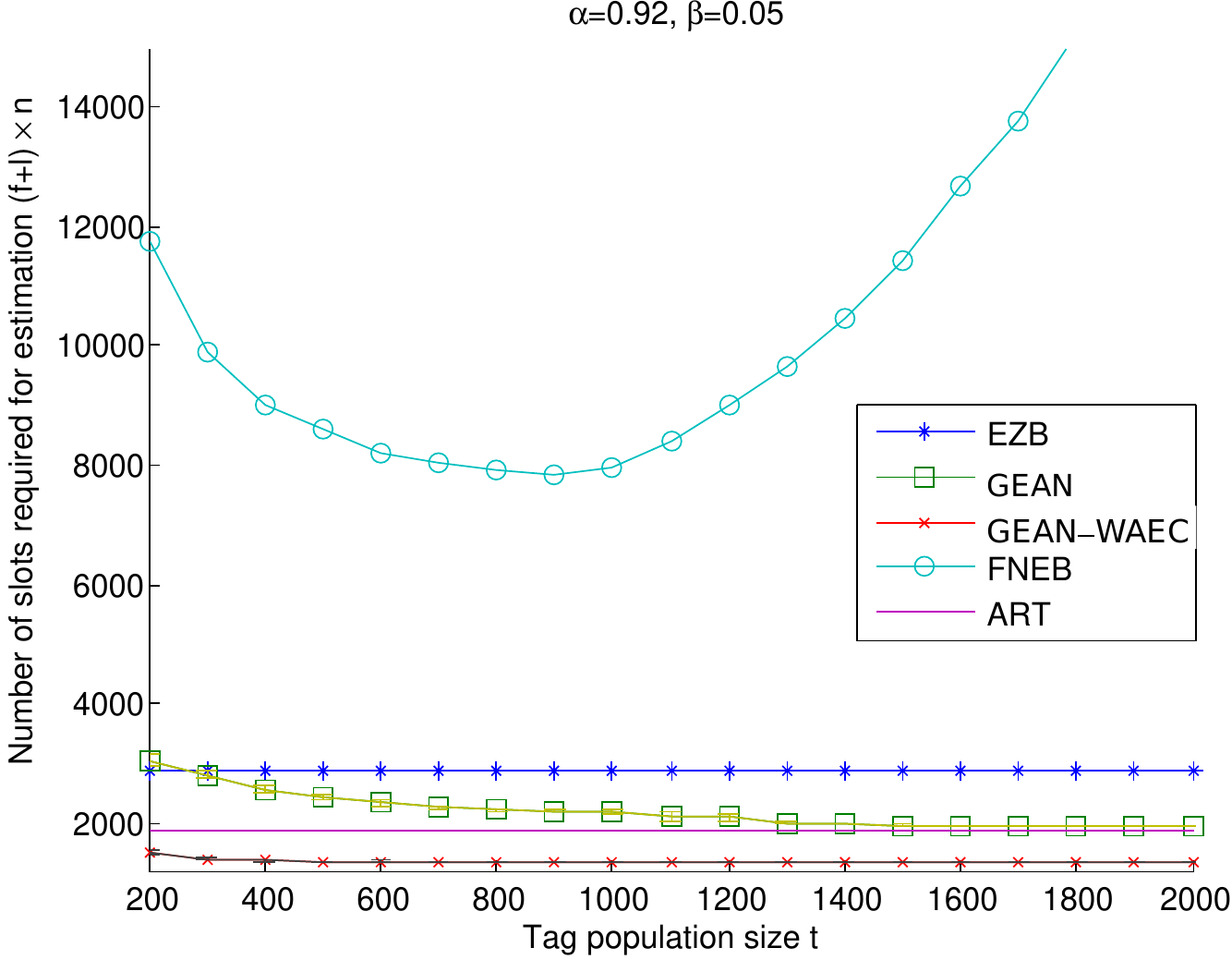}
\caption{Number of slots required by different tag estimation schemes against different tag population sizes  . $\alpha=0.92, \beta=0.05$}
\label{r01_f8}
\end{figure}

After the completion of the analytical analysis of GEAN we used MATLAB to get simulation results for GEAN. We compared the performance of GEAN with few other recent works. The comparison is divided into two chambers complementary to each other. In one chamber we have the acheieved reliability by different schemes and in the other one we have the cost the respective schemes have to pay in terms of the number of slots. To put things into perspective and ensure a fair comparison we presented GEAN both with and without the approximation error counted. In Figures \ref{r01_f5} and \ref{r01_f7} we see that for two different acuracy requirements GEAN and EZB achieve the required reliability and GEAN Without the Approximation Error Counted (GEAN-WAEC) falls a little short as expected. The remaining two schemes fall a good distance short in meeting the reliability requirements. Figures \ref{r01_f6} and \ref{r01_f8} show the required number of slots required by different schemes to achieve the reliabilities given in Figures \ref{r01_f5} and \ref{r01_f7} respectively. We see that if we ignore the approximation error like others, we require less number of slots for estimaiton than the other three schemes. Taking the approximation error into account GEAN requires more number of slots than ART, but delivers much better reliability than ART. On the other note, EZB and GEAN both deliver the promised reliability but GEAN outperforms EZB in terms of the number of slots required for estimation except for very small AG population sizes. 

  In Figures~\ref{r01_f6} and \ref{r01_f8}, since $ (f+l)\times n $ is random due to the randomness of $ t_{m} $, we presented the mean value line along with the standard deviation over $ 50 $ samples.  The   gradual descent of the GEAN curves, is due to the fact that for the smaller values of $ t $ we can not shoot for the bigger values of $ r $ due to restrictions incurred by the condition $ f_{max} \geq f_{man} $. As $ t $ gets bigger we can shoot for the larger values of $ r $ and Figure~\ref{v} suggests that the variance is smaller for the bigger values of $ r $. That saves us in terms of the number of rounds. The fact that $ (f+l)\times n $ curve eventually gets saturated, happens because   $ k(r) $ given by \eqref{kaka} has a very sharp increment for the bigger values of $ r $. This trend cost us in terms of the  frame size required to satisfy \eqref{33}. So, even if $ t $ continues increasing we cannot operate beyond a certain value of $ r $ for a given accuracy requirements, hence is the saturation.   
  
\begin{figure}
\centering
\includegraphics[scale=0.7]{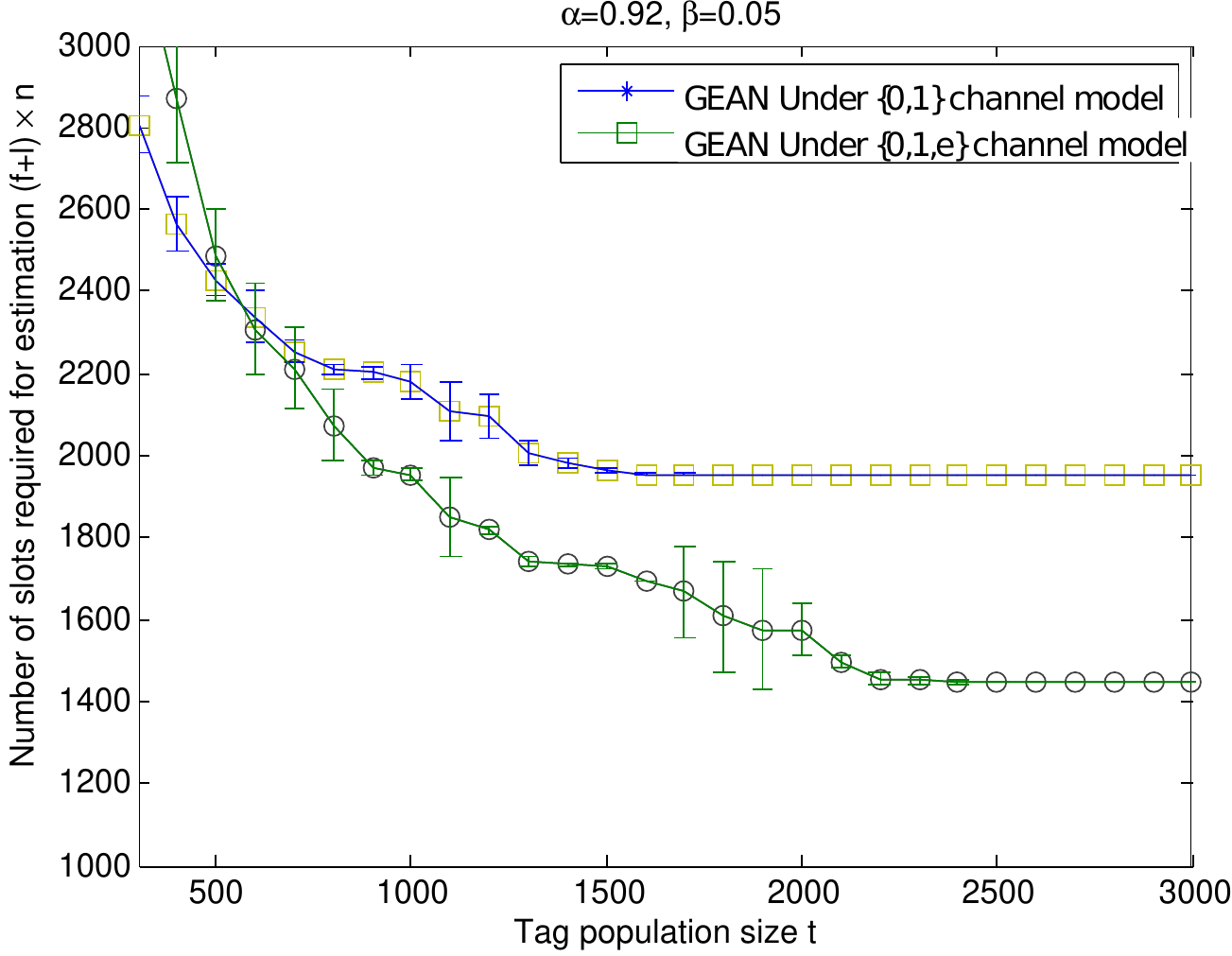}
\caption{Number of slots required by GEAN under two different channel models against the tag population size $ t $}
\label{comp}
\end{figure}

  As a side note we present the performance of GEAN under $ \{0,1,e\} $ channel model in Figure \ref{comp}.   Under $ \{0,1,e\} $   each $ 1 $ is  certain.  Because of this added certainty, we generally  need  fewer number of slots to meet the same  accuracy requirements under $ \{0,1,e\} $  than under $ \{0,1\} $ channel model. The only aberration noticable   is because under $ \{0,1,e\} $ channel model for smaller values of $ t $ we can not operate on the smaller values of $ r $ because of the singularity issue. That in turn costs in terms of the number of slots required. 



\section{Conclusion} 
The key contribution of this paper is that it proposes a completely new and more effective technique for the estimation of an AG population size that communicate with the same AP. The technique is primarily based on the idea of a big enough frame size and  precise approximation of the estimator to Gaussian distribution. These features provide us with advantages over other schemes both in terms of estimation accuracy and cost savings.  The simulated results clearly demostrate the rigorous analytical foundation of our work    under both $ \{0,1\} $ and $ \{0,1,e\} $  channel models.

\appendices
\section{Proof of Lemma~\ref{l1}}
\begin{proof}
Using equation~\eqref{17}, and the fact that for $x \ll 1$ and $y \gg 1$, $(1-x)^{y}$ can be appriximated as $e^{y\ln[1-x]}$ and that in turn can be reduced to $e^{-xy}$ applying Taylor series, we get,
 
\begin{align}\label{gft}
&g_{f}(t)=1-2\left( 1-\frac{p}{f•}\right)^{t}=1-2e^{-\frac{tp}{f}}=1-2e^{-r}\notag
\end{align}
\begin{align}
\Rightarrow & \frac{d}{dr•}[g_{f}(t)]= 2e^{-r}\notag\\
\Rightarrow &\frac{d^{2}}{dr^{2}•}[g_{f}(t)]= -2e^{-r}\notag
\end{align}
In our algorithm we have $ p\in (0,1] $, $ f\geq 1 $ and $ t\geq 1 $. So, the first and second derivatives of $ g_{f}(t) $ with respect to $ r $ will always be non-negative and negative respectively. Hence, $ g_{f}(t) $ is a monotonically increasing function of $ r $. 
\end{proof}

\section{Derivation of Lindeberg Feller conditions for GEAN under $ \{0,1\} $ channel model}
\subsubsection*{\textbf{First condition}}
From equation \eqref{a2} we have the following,
\begin{align}
&\left|1 - \mu_{f}\right|> \epsilon \sigma_{j,f} \notag\\
\Rightarrow &1 - \mu_{f} > \epsilon \sigma_{j,f}\notag\\
\Rightarrow &1 - (p_{n}- p_{0}) > \epsilon \sigma_{j,f}\notag\\
\Rightarrow & 1 - 1+2 p_{0}> \epsilon \sigma_{j,f}\notag\\
\Rightarrow & 2 p_{0}> \epsilon \sigma_{j,f}\notag\\
\Rightarrow & 4 p_{0}^{2}> \epsilon^{2} f \left[ p_{n} + p_{0}-( p_{n}- p_{0})^{2}\right]\notag\\
\Rightarrow & -kp_{n} -kp_{0} +p_{n}^{2}k -2 p_{n}p_{0}k +p_{0}^{2}k +4p_{0}^{2} > 0 \
\end{align}
Making the same approximation as we did in Appendix A, we can say \eqref{a2} does not hold if the following is satisfied,
\begin{align}
k\geq \left| \frac{e^{-r}}{1-e^{-r}} \right|=k_{1}=\frac{e^{-r}}{1-e^{-r}} 
\end{align}
\subsubsection*{\textbf{Second condition}}
From equation \eqref{a3} we have the following,
\begin{align}
&\left|-1 - \mu_{f}\right|> \epsilon \sigma_{j,f}\notag\\
\Rightarrow &1 + \mu_{f} > \epsilon \sigma_{j,f}\notag\\
\Rightarrow &1 + (p_{n}- p_{0}) > \epsilon \sigma_{j,f}\notag\\
\Rightarrow &1 + 1-2 p_{0}> \epsilon \sigma_{j,f}\notag\\
\Rightarrow &2- 2 p_{0}> \epsilon \sigma_{j,f}\notag\\
\Rightarrow &4- 8p_{0}+ 4 p_{0}^{2}> \epsilon^{2} f \left[ p_{n} + p_{0}-( p_{n}- p_{0})^{2}\right]\notag\\
\Rightarrow &4- 8p_{0}+ 4 p_{0}^{2} -kp_{n} -kp_{0} +p_{n}^{2}k -2 p_{n}p_{0}k +p_{0}^{2}k > 0 \label{k2}
\end{align}
Simple algebraic manipulations and aforementioned approximation gives us, \eqref{k2}  or equivalently \eqref{a3} does not hold if the following is satisfied,
\begin{align}
k\geq \left| \frac{1-e^{-r}}{e^{-r}} \right|=k_{2}=\frac{1-e^{-r}}{e^{-r}} 
\end{align}

From the above two conditions we see that if we select the value of $k$ such that $k=max\{k_{1}, k_{2}\}$,  \eqref{a2} and \eqref{a3}  will not hold or equivalently \eqref{lfc} will hold. That essentially means for $k=max\{k_{1}, k_{2}\}$ the GEAN estimator $Z_{f}$ under $ \{0,1\}$ channel model is Gaussian distributed. 

\section{Proof of  Lemma~\ref{lem1}}
\begin{proof}
To find the local minimum we need to differentiate the expectation curve and set the derivative to $0$. Solving that equation we will get the local minimum of the dip. Using \eqref{17},
\begin{align}
\frac{d}{dt}g_{f}(t) ] =   \frac{d}{dt}\left[1- \left(1-\frac{p}{f}\right)^{t}- 2 t\left(\frac{p}{f}\right)\left(1-\frac{p}{f}\right)^{t-1}\right]=0
\end{align}

Simple algebraic calculations give us, 

\begin{align}
t_{LM}=\frac{ -2 \left(\frac{p}{f}\right)  -   \left(1-\frac{p}{f}\right) \ln \left(1-\frac{p}{f}\right) } {2 \left(\frac{p}{f}\right)  \ln \left(1-\frac{p}{f}\right)}
\end{align}

Here, $t_{LM}$ stands for the $t$ value where the local minimum of the dip occurs (i.e. at point $D$ in Figure:\ref{e}). Now, since the value of $\frac{p}{f}<<1$  we can approximate $\ln \left(1-\frac{p}{f}\right)$ as $-\frac{p}{f}$. This will give us the following, $ t_{LM} \approx \frac{f}{2p}$. Which means the local minimum for the dip occurs at a value of $t=t_{LM}$ which is supported by our simulation results. Substituting, $t_{LM}$ in \eqref{r} gives us $r_{LM}=\frac{1}{2}$.
\end{proof}

\section{Proof of  Lemma~\ref{lem2}}
\begin{proof}
To find an inverse we need $g_{f}(t)$ to be a monotonic function of $t$ . To find which part of the $g_{f}(t)$ demonstrates monotonic behavior we need the second derivative of $g_{f}(t)$ and check for it's convexity and concavity characteristics. Again using \eqref{17}

\begin{align}
\begin{split}
\frac{d^{2}}{dt^{2}} g_{f}(t) &=\left(1-\frac{p}{f}\right)^{t-1} \ln \left(1-\frac{p}{f}\right) \left[ -2 t\left(\frac{p}{f}\right) \ln \left(1-\frac{p}{f}\right) \right. \\ & \left. -  4 \left(\frac{p}{f}\right) - \left(1-\frac{p}{f}\right) \ln \left(1-\frac{p}{f}\right)\right]
\end{split}
\end{align}

If we closely follow the equation we see that, the value of the common factor $ \left(1-\frac{p}{f}\right)^{t-1} \ln \left(1-\frac{p}{f}\right) $ is negative . So, for the total value to be posive the part of the equation inside the third bracket will have to be negative. 
After algebraic manipulations and  approximating $\ln \left(1-\frac{p}{f}\right)$ as $-\frac{p}{f}$ we get for $ t< \frac{3f+p}{2p}$,
\begin{align}
 \left[ 2 t\left(\frac{p}{f}\right) \left(\frac{p}{f}\right)     -4 \left(\frac{p}{f}\right) + \left(1-\frac{p}{f}\right)  \left(\frac{p}{f}\right)\right]<0 \notag
\end{align}
Or, equivalently, for $ t< \frac{3f+p}{2p} \approx  \frac{3f}{2p}$, $\frac{d^{2}}{dt^{2}}g_{f}(t) $ is positive, indicating $g_{f}(t) $ is a convex function of $t$ and for the rest of the $t$ values the curve is concave. Substituting $ t= \frac{3f}{2p}$ in \eqref{r} gives us $r=\frac{3}{2}$. 
\end{proof}

\section{Proof of  Lemma~\ref{lem3}}
\begin{proof}
At point $B$ in Figure~\ref{e}, we know the corresponding frame size is $f_{max}$, uning \eqref{r} and \eqref{mono} we have, 
\begin{align}\label{67}
\frac{t_{m}p}{f_{max}}\geq r_{min} \quad
\Rightarrow  &\frac{t_{m}p}{f_{max}}\geq 1.2564
\end{align}

Now from \cite{shahzad2015fast}, we know that the upper bound on $t$ is  always less that $2t$ for  $\alpha=90\%$  and $\beta = 10\%$ which are way less than the accuracy  level we are dealing with. For the tighter accuracy requirements the upperbound is even lower. Since $f_{max}\geq f$ using \eqref{67} we get, 
\begin{align}\label{69}
\frac{2tp}{f_{max}}\geq 1.2567 \quad \Rightarrow \frac{2tp}{f}\geq 1.2564 
\end{align}
Using \eqref{r} and \eqref{69} we get, $ r  \geq 0.6283>\frac{1}{2} $ 

\end{proof}
\section{Derivation of Lindeberg Feller conditions for GEAN under $ \{0,1,e\} $ channel model}
\subsubsection*{\textbf{First condition}}
From equation \eqref{30} we have the following,
\begin{align}
\Rightarrow &1- \mu_{f} > \epsilon \sqrt{f}\sigma_{j,f} \notag\\
\Rightarrow &1 - (p_{e} - p_{1}) >  \epsilon \sqrt { f   \left[p_{e} + p_{1}  -(p_{e} - p_{1})^{2}\right]} \notag\\
\Rightarrow &1 -  2(p_{e} -  p_{1})  + (p_{e} - p_{1})^{2} >  \epsilon^{2}f    \left[p_{e} + p_{1}  -    (p_{e} - p_{1})^{2}\right] \notag\\
\Rightarrow &1 - \left (2+ \epsilon^{2}f\right) p_{e}  + \left (2- \epsilon^{2}f\right)p_{1} + \left (1+\epsilon^{2}f\right) (p_{e} - p_{1})^{2}>0
\end{align}
Letting $\epsilon^{2}f $ be represented by $k$, and inserting the expression for $p_{0}$ , $p_{1}$  and $p_{e}$ we have, 
\begin{align}
 \Rightarrow &1 -    \left (2+ k\right) \left[  1-  \left(1-\frac{p}{f}\right)^{t}-  t\left(\frac{p}{f}\right)\left(1 -\frac{p}{f}\right)^{t-1}   \right]\notag\\&+ \left (2   - k\right) \left[   t\left(\frac{p}{f}\right)\left(1-\frac{p}{f}\right)^{t-1}  \right] +\left (1+k\right)  \left[  1+  \right. \notag\\ & \left. \left(1-\frac{p}{f}\right)^{2t}   + 4t^{2}\left(\frac{p}{f}\right)^{2} \left(1-\frac{p}{f}\right)^{2(t-1)}    - 2\left(1-\frac{p}{f}\right)^{t} \right. \notag\\ & \left.+  4t\left(\frac{p}{f}\right)  \left(1-\frac{p}{f}\right)^{2t-1} -   4t\left(\frac{p}{f}\right)\left(1-\frac{p}{f}\right)^{t-1}   \right] >0
\end{align}
We know, for $x \ll 1$ and $y \gg 1$, $(1-x)^{y}$ can be appriximated as $e^{y\ln[1-x]}$ and that in turn can be reduced to $e^{y[-x-\frac{1}{2} x^{2}]}$ applying Taylor series. Applying this we get,
\begin{align}
\Rightarrow &1-  \left (2+ k\right)  \left[  1- e^{-t\{\frac{p}{f} + \frac{1}{2}(\frac{p}{f})^{2}\}}     -  t\left(\frac{p}{f}\right) e^{-(t-1)\{\frac{p}{f} + \frac{1}{2}(\frac{p}{f})^{2}\}}  \right]  \notag\\ &+ \left (2- k\right)  \left[  t\left(\frac{p}{f}\right) e^{-(t-1)\{\frac{p}{f} + \frac{1}{2}(\frac{p}{f})^{2}\}}  \right]  
+  \left (1+k\right)\notag\\ &  \left[  1+  e^{-2t\{\frac{p}{f} + \frac{1}{2}(\frac{p}{f})^{2}\}} + 4t^{2}\left(\frac{p}{f}\right)^{2}  e^{-2(t-1)\{\frac{p}{f} + \frac{1}{2}(\frac{p}{f})^{2}\}}   \right. \notag\\ & \left. -   2 e^{-t\{\frac{p}{f} + \frac{1}{2}(\frac{p}{f})^{2}\}} \right]
 +   \left (1+k\right)\left [ 4t\left(\frac{p}{f}\right) e^{-(2t-1)\{\frac{p}{f} + \frac{1}{2}(\frac{p}{f})^{2}\}}  \right. \notag\\ & \left.- 4t\left(\frac{p}{f}\right) e^{-(t-1)\{\frac{p}{f} + \frac{1}{2}(\frac{p}{f})^{2}\}}   \right] > 0 
\end{align}
Now we have a list of approximations to make. They are,
\begin{align}
&e^{-t\{\frac{p}{f} + \frac{1}{2}(\frac{p}{f})^{2}\}} \approx e^{-t\frac{p}{f}}\label{59}\\
& e^{-(2t-1)\{\frac{p}{f} + \frac{1}{2}(\frac{p}{f})^{2}\}}\approx e^{-2t\frac{p}{f}} \label{60}\\
& e^{-2(t-1)\{\frac{p}{f} + \frac{1}{2}(\frac{p}{f})^{2}\}}\approx e^{-2t\frac{p}{f}} \label{61}\l\\
& e^{-(t-1)\{\frac{p}{f} + \frac{1}{2}(\frac{p}{f})^{2}\}}\approx e^{-t\frac{p}{f}}\label{62}
\end{align}

After all these approximations and using \eqref{r} we have,
%
%
%
%
%
\begin{align}\label{63}
\Rightarrow &1-  \left (2+ k\right)  \left(  1- e^{-r}    - r e^{-r}  \right)   + \left (2- k\right)   r  e^{-r} +\left (1+k\right)  \left( 1+\right. \notag\\ & \left.  e^{-2r} + 4r^{2}   e^{-2r}  - 2e^{-r}\right)
  +\left (1+k\right)\left ( 4r e^{-2r} - 4r e^{-r}   \right) > 0  \notag\\
\end{align}
For the equation \eqref{30} to not hold, \eqref{63} must not hold. Simple algebraic manipulations give us that for \eqref{63} to not hold the value of $k$ must be, 
%
%
%
%
%
%
\begin{align}
k \geq \frac{1}{\left|\frac{-e^{r} \left( 1 +4r\right)}{\left(1+ 2r \right)^{2}}\right | -1}=k_{1} \notag
\end{align}
So, $k_{1}$ is the minimum value of $k$ for which \eqref{30} does not hold. 
\subsubsection*{\textbf{2nd condition}}
From equation \eqref{31} we have the following,
\begin{align}
 \Rightarrow &1+ \mu_{f} > \epsilon  \sqrt{f}\sigma_{j,f}\notag\\
\Rightarrow &1 + (p_{e} - p_{1}) >  \epsilon \sqrt { f   \left[p_{e} + p_{1}  -(p_{e} - p_{1})^{2}\right]} \notag\\
\Rightarrow &1   + 2(p_{e} -  p_{1})  + (p_{e} - p_{1})^{2} >  \epsilon^{2}f    \left[P_{e} + p_{1}  -(p_{e} - p_{1})^{2}\right] \notag\\
\Rightarrow &1 +    \left (2- \epsilon^{2}f\right) p_{e}     - \left (2+ \epsilon^{2}f\right)p_{1} +\left (1+\epsilon^{2}f\right) (p_{e} - p_{1})^{2} >  0
\end{align}
letting $\epsilon^{2}f $ be represented by $k$, and inserting the expression for $p_{0}$ , $p_{1}$  and $p_{e}$ we have, 
\begin{align}
\Rightarrow &1 +    \left (2- k\right) \left[  1-  \left(1-\frac{p}{f}\right)^{t}-  t\left(\frac{p}{f}\right)\left(1-\frac{p}{f}\right)^{t-1}   \right]    - \left (2+  k\right)\notag\\ & \left[   t\left(\frac{p}{f}\right)\left(1-\frac{p}{f}\right)^{t-1}  \right] +\left (1+k\right) 
   \left[  1+  \left(1-\frac{p}{f}\right)^{2t}  + 4t^{2}\left(\frac{p}{f}\right)^{2}\right. \notag\\ & \left. \left(1-\frac{p}{f}\right)^{2(t-1)}    - 2\left(1-\frac{p}{f}\right)^{t} +  4t\left(\frac{p}{f}\right) \left(1-\frac{p}{f}\right)^{2t-1}  \right.  \notag\\ & \left. - 4t\left(\frac{p}{f}\right)\left(1-\frac{p}{f}\right)^{t-1}   \right] >0
\end{align}
Like the first condition $(1-x)^{y}$ can be appriximated as $e^{y\ln[1-x]}$ and that in turn can be reduced to $e^{y[-x-\frac{1}{2} x^{2}]}$ applying Taylor series. Applying this along with   approximations made in \eqref{59}, \eqref{60}, \eqref{61}, \eqref{62} and using \eqref{r} we have, 
%
%
%
%
%
%
\begin{align}\label{66}
\Rightarrow &1+  \left (2- k\right)  \left(  1- e^{-r}     -  r e^{-r}  \right)   - \left (2+ k\right)  r e^{-r} +\left (1+k\right) \notag\\ &   \left( 1+  e^{-2r} + 4r^{2}   e^{-2r}   - 2e^{-r}\right)
 +\left (1+k\right)\left ( 4re^{-2r} - 4re^{-r}   \right)> 0  
\end{align}
For the equation \eqref{31} to not hold, \eqref{66} must not hold. Simple algebraic manipulations give us that for \eqref{66} to not hold the value of $k$ must be, 
\begin{align}
k \geq \left|\frac { e^{2r} +  \left(1+2r \right) \left(  \frac{1}{4} - e^{r} \right)}{   \frac{1}{4}  \left(1+2 r\right)^{2}   - r e^{r}}\right|= k_{2}  \notag
\end{align}
So, $k_{2}$ is the minimum value of $k$ for which \eqref{31} does not hold. 
\subsubsection*{\textbf{3rd condition}}
From equation \eqref{32} we have the following,
\begin{align}
 \Rightarrow &\mu_{f} > \epsilon \sqrt{f}\sigma_{j,f}  \notag\\
 \Rightarrow &(p_{e} - p_{1}) >  \epsilon \sqrt { f   \left[p_{e} + p_{1}  -(p_{e} - p_{1})^{2}\right]} \notag\\
 \Rightarrow &(p_{e} - p_{1})^{2} >  \epsilon^{2}f    \left[p_{e} + p_{1}  -(p_{e} - p_{1})^{2}\right] \notag\\
\Rightarrow &- \epsilon^{2}f( p_{e} + p_{1} )+\left (1+\epsilon^{2}f\right) (p_{e} - p_{1})^{2} >  0  \notag\\
\Rightarrow &- k \left[  1-  \left(1-\frac{p}{f}\right)^{t}  \right]     +\left (1+k\right) 
\left[  1+  \left(1-\frac{p}{f}\right)^{2t} + 4t^{2} \left(\frac{p}{f}\right)^{2} \right. \notag\\ & \left. \left(1-\frac{p}{f}\right)^{2(t-1)}    - 2\left(1-\frac{p}{f}\right)^{t} +   4t\left(\frac{p}{f}\right)  \left(1-\frac{p}{f}\right)^{2t-1} \right. \notag\\ & \left.- 4t\left(\frac{p}{f}\right)\left(1-\frac{p}{f}\right)^{t-1}   \right] >0
\end{align}

Like we did in the previous two conditions, $(1-x)^{y}$ can be appriximated as $e^{y\ln[1-x]}$ and that in turn can be reduced to $e^{y[-x-\frac{1}{2} x^{2}]}$ applying Taylor series. Applying this along with   approximations made in \eqref{59}, \eqref{60}, \eqref{61}, \eqref{62} and using \eqref{r}  we have, 
%
%
%
%
%
%
\begin{align}\label{68}
\Rightarrow &- k \left( 1- e^{-r} \right)  +\left (1+k\right)    \left(  1+  e^{-2r} + 4r^{2}   e^{-2r} - 2e^{-r}\right) \notag\\ &+\left (1+k\right)\left ( 4re^{-2r} - 4r e^{-r}   \right) > 0 
\end{align}

For the equation \eqref{32} to not hold, \eqref{68} must not hold. Simple algebraic manipulations give us that for \eqref{68} to not hold the value of $k$ must be, 
%
%
%
%
%
%
\begin{align}
k \geq \left|\frac{e^{2r} -2e^{r} \left(1+ 2 r \right) + \left(1+2r\right)^{2}}{\left(1+2r\right)^{2}  -e^{r} \left(1 +4 r \right)}\right|=k_{3}  \notag
\end{align}

So, $k_{3}$ is the minimum value of $k$ for which \eqref{32} does not hold. 

From the above three conditions we see that if we select the value of $k$ such that $k=max\{k_{1}, k_{2}, k_{3}\}$  all of \eqref{30}, \eqref{31} and \eqref{32} will not hold or equivalently \eqref{lfc} will hold. That essentially means for $k=max\{k_{1}, k_{2}, k_{3}\}$ the GEAN estimator $Z_{f}$ under $ \{0,1,e\}$ channel model  is Gaussian distributed.

\bibliography{ref}

\begin{thebibliography}{10}
\providecommand{\url}[1]{#1}
\csname url@samestyle\endcsname
\providecommand{\newblock}{\relax}
\providecommand{\bibinfo}[2]{#2}
\providecommand{\BIBentrySTDinterwordspacing}{\spaceskip=0pt\relax}
\providecommand{\BIBentryALTinterwordstretchfactor}{4}
\providecommand{\BIBentryALTinterwordspacing}{\spaceskip=\fontdimen2\font plus
\BIBentryALTinterwordstretchfactor\fontdimen3\font minus
  \fontdimen4\font\relax}
\providecommand{\BIBforeignlanguage}[2]{{%
\expandafter\ifx\csname l@#1\endcsname\relax
\typeout{** WARNING: IEEEtran.bst: No hyphenation pattern has been}%
\typeout{** loaded for the language `#1'. Using the pattern for}%
\typeout{** the default language instead.}%
\else
\language=\csname l@#1\endcsname
\fi
#2}}
\providecommand{\BIBdecl}{\relax}
\BIBdecl

\bibitem{nemmaluri2008sherlock}
A.~Nemmaluri, M.~D. Corner, and P.~Shenoy, ``Sherlock: automatically locating
  objects for humans,'' in \emph{Proceedings of the 6th international
  conference on Mobile systems, applications, and services}.\hskip 1em plus
  0.5em minus 0.4em\relax ACM, 2008, pp. 187--198.

\bibitem{lee2008efficient}
C.-H. Lee and C.-W. Chung, ``Efficient storage scheme and query processing for
  supply chain management using {RFID},'' in \emph{Proceedings of the 2008 ACM
  SIGMOD international conference on Management of data}.\hskip 1em plus 0.5em
  minus 0.4em\relax ACM, 2008, pp. 291--302.

\bibitem{klaus2010rfid}
F.~Klaus, ``Rfid handbook: Fundamentals and applications in contactless smart
  cards, radio frequency identification and near-field communication,'' 2010.

\bibitem{zhao2019sensor}
Y.~Zhao, Z.~Li, B.~Hao, and J.~Shi, ``Sensor selection for tdoa-based
  localization in wireless sensor networks with non-line-of-sight condition,''
  \emph{IEEE Transactions on Vehicular Technology}, vol.~68, no.~10, pp.
  9935--9950, 2019.

\bibitem{zhang2019heterogeneous}
Y.~Zhang, R.~Wang, M.~S. Hossain, M.~F. Alhamid, and M.~Guizani,
  ``Heterogeneous information network-based content caching in the internet of
  vehicles,'' \emph{IEEE Transactions on Vehicular Technology}, vol.~68,
  no.~10, pp. 10\,216--10\,226, 2019.

\bibitem{gurugopinath2019cache}
S.~Gurugopinath, P.~C. Sofotasios, Y.~Al-Hammadi, and S.~Muhaidat,
  ``Cache-aided non-orthogonal multiple access for 5g-enabled vehicular
  networks,'' \emph{arXiv preprint arXiv:1906.06025}, 2019.

\bibitem{zhang2019performance}
M.~Zhang, P.~H.~J. Chong, and B.-C. Seet, ``Performance analysis and boost for
  a mac protocol in vehicular networks,'' \emph{IEEE Transactions on Vehicular
  Technology}, vol.~68, no.~9, pp. 8721--8728, 2019.

\bibitem{kodialam2006fast}
M.~Kodialam and T.~Nandagopal, ``Fast and reliable estimation schemes in {RFID}
  systems,'' in \emph{Proceedings of the 12th annual international conference
  on Mobile computing and networking}.\hskip 1em plus 0.5em minus 0.4em\relax
  ACM, 2006, pp. 322--333.

\bibitem{kodialam2007anonymous}
M.~Kodialam, T.~Nandagopal, and W.~C. Lau, ``Anonymous tracking using {RFID}
  tags,'' in \emph{IEEE INFOCOM 2007-26th IEEE International Conference on
  Computer Communications}.\hskip 1em plus 0.5em minus 0.4em\relax IEEE, 2007,
  pp. 1217--1225.

\bibitem{han2010counting}
H.~Han, B.~Sheng, C.~C. Tan, Q.~Li, W.~Mao, and S.~Lu, ``Counting {RFID} tags
  efficiently and anonymously,'' in \emph{INFOCOM, 2010 Proceedings
  IEEE}.\hskip 1em plus 0.5em minus 0.4em\relax IEEE, 2010, pp. 1--9.

\bibitem{li2010energy}
T.~Li, S.~Wu, S.~Chen, and M.~Yang, ``Energy efficient algorithms for the
  {RFID} estimation problem,'' in \emph{INFOCOM, 2010 Proceedings IEEE}.\hskip
  1em plus 0.5em minus 0.4em\relax IEEE, 2010, pp. 1--9.

\bibitem{shah2009anonymous}
V.~Shah-Mansouri and V.~W. Wong, ``Anonymous cardinality estimation in {RFID}
  systems with multiple readers,'' in \emph{Global Telecommunications
  Conference, 2009. GLOBECOM 2009. IEEE}.\hskip 1em plus 0.5em minus
  0.4em\relax IEEE, 2009, pp. 1--6.

\bibitem{zanella2012estimating}
A.~Zanella, ``Estimating collision set size in framed slotted aloha wireless
  networks and {RFID} systems,'' \emph{IEEE Communications Letters}, vol.~16,
  no.~3, pp. 300--303, 2012.

\bibitem{shah2011cardinality}
V.~Shah-Mansouri and V.~W. Wong, ``Cardinality estimation in {RFID} systems
  with multiple readers,'' \emph{IEEE Transactions on Wireless Communications},
  vol.~10, no.~5, pp. 1458--1469, 2011.

\bibitem{shahzad2015fast}
M.~Shahzad and A.~X. Liu, ``Fast and accurate estimation of {RFID} tags,''
  \emph{IEEE/ACM Transactions on Networking}, vol.~23, no.~1, pp. 241--254,
  2015.

\bibitem{shahzad2013probabilistic}
------, ``Probabilistic optimal tree hopping for rfid identification,''
  \emph{ACM SIGMETRICS Performance Evaluation Review}, vol.~41, no.~1, pp.
  293--304, 2013.

\bibitem{araujo1980central}
A.~Araujo and E.~Gin{\'e}, \emph{The central limit theorem for real and Banach
  valued random variables}.\hskip 1em plus 0.5em minus 0.4em\relax Wiley New
  York, 1980, vol. 431.

\bibitem{hasan2018estimation}
M.~M. Hasan, S.~Wei, and R.~Vaidyanathan, ``Estimation of {RFID} tag population
  size by {Gaussian} estimator,'' in \emph{2018 IEEE International Conference
  on Communications (ICC)}.\hskip 1em plus 0.5em minus 0.4em\relax IEEE, 2018,
  pp. 1--6.

\bibitem{epcglobal2004radio}
E.~EPCglobal, ``Radio-frequency identity protocols class-1 generation-2 uhf
  {RFID} protocol for communications at 860 mhz--960 mhz version 1.0. 9,''
  \emph{K. Chiew et al./On False Authenticationsfor C1G2 Passive RFID Tags},
  vol.~65, 2004.

\bibitem{flajolet1985probabilistic}
P.~Flajolet and G.~N. Martin, ``Probabilistic counting algorithms for data base
  applications,'' \emph{Journal of computer and system sciences}, vol.~31,
  no.~2, pp. 182--209, 1985.

\bibitem{reza2011rfid}
A.~W. Reza, T.~K. Geok, K.~J. Chia, and K.~Dimyati, ``{RFID} transponder
  collision control algorithm,'' \emph{Wireless Personal Communications},
  vol.~59, no.~4, pp. 689--711, 2011.

\bibitem{alfian2019false}
G.~Alfian, M.~Syafrudin, B.~Yoon, and J.~Rhee, ``False positive rfid detection
  using classification models,'' \emph{Applied Sciences}, vol.~9, no.~6, p.
  1154, 2019.

\bibitem{massawe2012reducing}
L.~V. Massawe, J.~D. Kinyua, and H.~Vermaak, ``Reducing false negative reads in
  rfid data streams using an adaptive sliding-window approach,''
  \emph{Sensors}, vol.~12, no.~4, pp. 4187--4212, 2012.

\bibitem{vogt2002efficient}
H.~Vogt, ``Efficient object identification with passive {RFID} tags,'' in
  \emph{International Conference on Pervasive Computing}.\hskip 1em plus 0.5em
  minus 0.4em\relax Springer, 2002, pp. 98--113.

\bibitem{cha2005novel}
J.-R. Cha and J.-H. Kim, ``Novel anti-collision algorithms for fast object
  identification in {RFID} system,'' in \emph{11th International Conference on
  Parallel and Distributed Systems (ICPADS'05)}, vol.~2.\hskip 1em plus 0.5em
  minus 0.4em\relax IEEE, 2005, pp. 63--67.

\bibitem{brown1971martingale}
B.~M. Brown \emph{et~al.}, ``Martingale central limit theorems,'' \emph{The
  Annals of Mathematical Statistics}, vol.~42, no.~1, pp. 59--66, 1971.

\bibitem{qian2011cardinality}
C.~Qian, H.~Ngan, Y.~Liu, and L.~M. Ni, ``Cardinality estimation for
  large-scale {RFID} systems,'' \emph{IEEE Transactions on Parallel and
  Distributed Systems}, vol.~22, no.~9, pp. 1441--1454, 2011.

\end{thebibliography}
\bibliographystyle{IEEEtran}

\end{document}